\documentclass[apj,iop]{emulateapj}  
\usepackage{graphicx} 
\usepackage{apjfonts} 
\usepackage{hyperref}

\shorttitle{Chromospheric heating due to cancellation of quiet Sun internetwork fields} 

\shortauthors{Go\v{s}i\'{c} et al.}

\begin{document}

\title{Chromospheric heating due to cancellation of quiet Sun internetwork fields}
\author{M.~Go\v{s}i\'{c}}
\affil{Lockheed Martin Solar and Astrophysics Laboratory, Palo Alto, CA 94304, USA; mgosic@lmsal.com}
\affil{Bay Area Environmental Research Institute, Moffett Field, CA 94035, USA}

\author{J.~de la Cruz Rodr\'iguez}
\affil{Institute for Solar Physics, Dept. of Astronomy, Stockholm University, AlbaNova University Centre, SE-106 91, Stockholm, Sweden}

\author{B.~De Pontieu}
\affil{Lockheed Martin Solar and Astrophysics Laboratory, Palo Alto, CA 94304, USA; mgosic@lmsal.com}
\affil{Institute of Theoretical Astrophysics, University of Oslo, P.O. Box 1029 Blindern, NO-0315 Oslo, Norway}
\affil{Rosseland Centre for Solar Physics, University of Oslo, P.O. Box 1029 Blindern, NO-0315 Oslo, Norway}

\author{L.~R.~Bellot Rubio}
\affil{Instituto de Astrof\'{\i}sica de Andaluc\'{\i}a (IAA-CSIC),
	Apdo.\ 3004, 18080 Granada, Spain} 

\author{M. Carlsson}
\affil{Institute of Theoretical Astrophysics, University of Oslo, P.O. Box 1029 Blindern, NO-0315 Oslo, Norway}
\affil{Rosseland Centre for Solar Physics, University of Oslo, P.O. Box 1029 Blindern, NO-0315 Oslo, Norway}

\author{S.~Esteban Pozuelo}
\affil{Institute for Solar Physics, Dept. of Astronomy, Stockholm University, AlbaNova University Centre, SE-106 91, Stockholm, Sweden}
\affil{Instituto de Astrof\'{\i}sica de Andaluc\'{\i}a (IAA-CSIC),
	Apdo.\ 3004, 18080 Granada, Spain} 

\author{A.~Ortiz}
\affil{Institute of Theoretical Astrophysics, University of Oslo, P.O. Box 1029 Blindern, NO-0315 Oslo, Norway}
\affil{Rosseland Centre for Solar Physics, University of Oslo, P.O. Box 1029 Blindern, NO-0315 Oslo, Norway}
\affil{Instituto de Astrof\'{\i}sica de Andaluc\'{\i}a (IAA-CSIC),
	Apdo.\ 3004, 18080 Granada, Spain} 

\author{V.~Polito}
\affil{Smithsonian Astrophysical Observatory, 60 Garden Street, MS 58, Cambridge, MA 02138, USA} 

\begin{abstract}

The heating of the solar chromosphere remains one of the most important questions in solar physics. Our current understanding is that small-scale internetwork (IN) magnetic fields play an important role as a heating agent. Indeed, cancellations of IN magnetic elements in the photosphere can produce transient brightenings in the chromosphere and transition region. These bright structures might be the signature of energy release and plasma heating, probably driven by magnetic reconnection of IN field lines. Although single events are not expected to release large amounts of energy, their global contribution to the chromosphere may be significant due to their ubiquitous presence in quiet Sun regions. In this paper we study cancellations of IN elements and analyze their impact on the energetics and dynamics of the quiet Sun atmosphere. We use high resolution, multiwavelength, coordinated observations obtained with the Interface Region Imaging Spectrograph (IRIS) and the Swedish 1-m Solar Telescope (SST) to identify cancellations of IN magnetic flux patches and follow their evolution. We find that, on average, these events live for $\sim$$3$~minutes in the photosphere and $\sim$$12$~minutes in the chromosphere and/or transition region. Employing multi-line inversions of the \ion{Mg}{2} h \& k lines we show that cancellations produce clear signatures of heating in the upper atmospheric layers. However, at the resolution and sensitivity accessible to the SST, their number density still seems to be one order of magnitude too low to explain the global chromospheric heating.
\end{abstract}


\section{Introduction}

Outside of sunspots and active regions, the so-called quiet Sun (QS) is pervaded by strong kG fields located at the boundaries of supergranular cells---photospheric network (NE). In between the NE there are small and highly transient internetwork (IN) fields.

Recent results by \cite{Gosicetal2016} proposed IN fields as essential contributors to the solar magnetism \citep[see also][]{TrujilloBueno2004}. The authors showed that IN elements bring magnetic flux to the solar surface at an enormous rate of 120~Mx~cm$^{-2}$~day$^{-1}$, much higher than active regions \citep[1~Mx~cm$^{-2}$~day$^{-1}$;][]{ThorntonParnell}. Part of that flux is dragged by convective motions toward the closest intergranular lanes \citep[see, e.g.,][]{MartinezGonzalezBellotRubio} and then toward the NE \citep{LivingstonHarvey, Zirin1985, WangZirin, Orozco2012}, providing as much unsigned magnetic flux as is present in the NE in only about 10 hours \citep{Gosicetal2014}. The rest of the flux disappears through in situ fading or cancels with opposite polarity IN patches before reaching the NE. According to \cite{Gosicetal2016}, fading and IN flux transfer to the NE are the dominant flux removal processes in supergranular cells. However, being ubiquitous, small-scale cancellations driven by granular and supergranular flows may hold the key to decipher a fundamental problem in solar physics, namely, chromospheric and coronal heating \citep[e.g.,][]{LongcopeKankelborg, Priestetal2002, GalsgaardParnell}. Therefore, it is important to understand how cancellation of IN elements affects the upper solar atmosphere.

Cancellation of magnetic flux can take place at the junction of intergranular \citep{Kuboetal2010} and mesogranular lanes (see Figure 7 in \citealt{Requereyetal2016}), and at the borders of supergranular cells where persistent sinks are observed \citep{Requereyetal2018}. It occurs when two opposite-polarity features come into close proximity \citep{Livietal1985, Martin1988}, leading to flux removal from the solar surface either by submergence of an $\Omega$--shaped loop below the photosphere or by the ascent of a $\text{U}$--shaped loop into the chromosphere (see Figure 2 in \citealt{Zwaan1987}). In both cases, the observational signatures in longitudinal magnetograms are the same: two opposite polarity patches approach each other, decreasing in size and strength, and eventually disappear below the detection limit. When the two canceling patches are part of the same magnetic system, the cancellation represents the submergence of a magnetic loop below the surface. This process is known as flux retraction. If the canceling patches were not previously connected by field lines (they are not part of the same system), the cancellation is the result of magnetic reconnection \citep{Zwaan1987, Priest1987}. When reconnection occurs below the surface, the opposite-polarity patches get connected by emerging U-shaped loops \citep{Spruitetal1987} and no thermal effects can be observed. If it happens above the surface, the patches get connected by a submerging $\Omega$ loop and the energy released may produce a local brightening in the upper atmosphere.

Using longitudinal magnetograms from Big Bear Solar Observatory, \cite{Zhangetal1998} estimated from lifetimes and cancellation rate of IN elements that the total energy released through IN cancellations is $\sim$$2\times10^{5}$~erg~cm$^{-2}$~s$^{-1}$, which is comparable to the energy required to heat the corona \citep[$3\times10^{5}$~erg~cm$^{-2}$~s$^{-1}$;][]{WithbroeNoyes}. This result is supported by \cite{Zhouetal2010} who used the same method in a high-resolution magnetogram sequence recorded with the Narrowband Filter Imager \citep[NFI;][]{Tsuneta} aboard the Hinode satellite \citep{2007SoPh..243....3K}. Recently, \cite{Meyeretal2013} used nonlinear force-free field extrapolations driven by a time series of magnetograms taken with the Helioseismic and Magnetic Imager
\citep[HMI;][]{Scherreretal2012} onboard the Solar Dynamics Observatory \citep[SDO;][]{Pesnelletal2012}, to study the build-up, storage and dissipation of the magnetic energy in the QS (IN+NE) corona. They concluded that the rate of energy dissipation (lower limit of $8.7\times10^{4}$~erg~cm$^{-2}$~s$^{-1}$) is roughly in agreement with the radiative losses of the QS corona. However, it is unclear whether any significant atmospheric heating (let alone coronal heating) is associated with IN cancellations, which is one of the reasons we undertake the current study.

On the other hand, \cite{Wiegelmannetal2013} investigated the temporal evolution of the magnetic connectivity in the QS and concluded that the energy released by reconnection processes is not sufficient to heat the chromosphere and corona. To that purpose, the authors used high temporal and spatial resolution observations from the Imaging Magnetograph eXperiment \citep[IMaX;][]{MartinezPilletetal2011} on board the \textsc{Sunrise} balloon borne observatory \citep{Solankietal2010, Bartholetal2011} to extrapolate magnetic fields above the photosphere, under the potential field approximation. \cite{Chittaetal2014} came to the same conclusion. These authors carried out magneto-frictional extrapolations of QS magnetic fields and argued that reconnections between QS magnetic elements cannot sustain the chromospheric and coronal heating.

The discrepancy between these results is likely due to the different instruments, techniques, and assumptions under which magnetic field extrapolations are carried out. For example, from magnetogram measurements only, it is not possible to know whether IN cancellations seen in the photosphere have an impact on the upper solar atmosphere. Regarding magnetic extrapolation techniques, potential (current-free) field studies provide a general idea of the global connectivity between magnetic elements, but neglect the possibility that QS magnetic fields may have a significant non-potential component \citep{WoodardChae1999, Zhaoetal2009}. Nonlinear force-free field extrapolations, such as the magnetofrictional method, are affected by boundary conditions. Depending on how magnetograms are preprocessed, the contribution from small-scale IN magnetic fields may completely be lost. Furthermore, there are indications that QS coronal magnetic fields are not force-free \citep{SchrijvervanBallegooijen2005}, therefore results obtained by magnetofrictional modelling of the quiet-Sun must be analyzed very carefully, as noted by \cite{Chittaetal2014}.

To determine whether IN magnetic fields can provide sufficient energy to maintain the chromospheric and coronal heating, it is essential to perform simultaneous, multi-instrument, multi-wavelength observations of IN regions with the highest sensitivity and resolution possible. To achieve this goal, in this paper we use space observations obtained with the Interface Region Imaging Spectrograph \citep[IRIS;][]{DePontieuetal2014}, accompanied by ground-based observations acquired with the Swedish 1-m Solar Telescope \citep[SST;][]{Scharmeretal2003}. These instruments give us the opportunity to study the evolution of the weak and small IN fields at high spatial, spectral, and temporal resolution, from the photosphere up to the transition region. We track canceling IN magnetic elements in the photosphere and examine their effects on the upper atmospheric layers. From imaging and spectroscopic measurements we will detect signatures of chromospheric heating and estimate the total energy released through cancellations of IN elements. 

\begin{figure*}[!t]
	\begin{center}
		\resizebox{1\hsize}{!}{\includegraphics[]{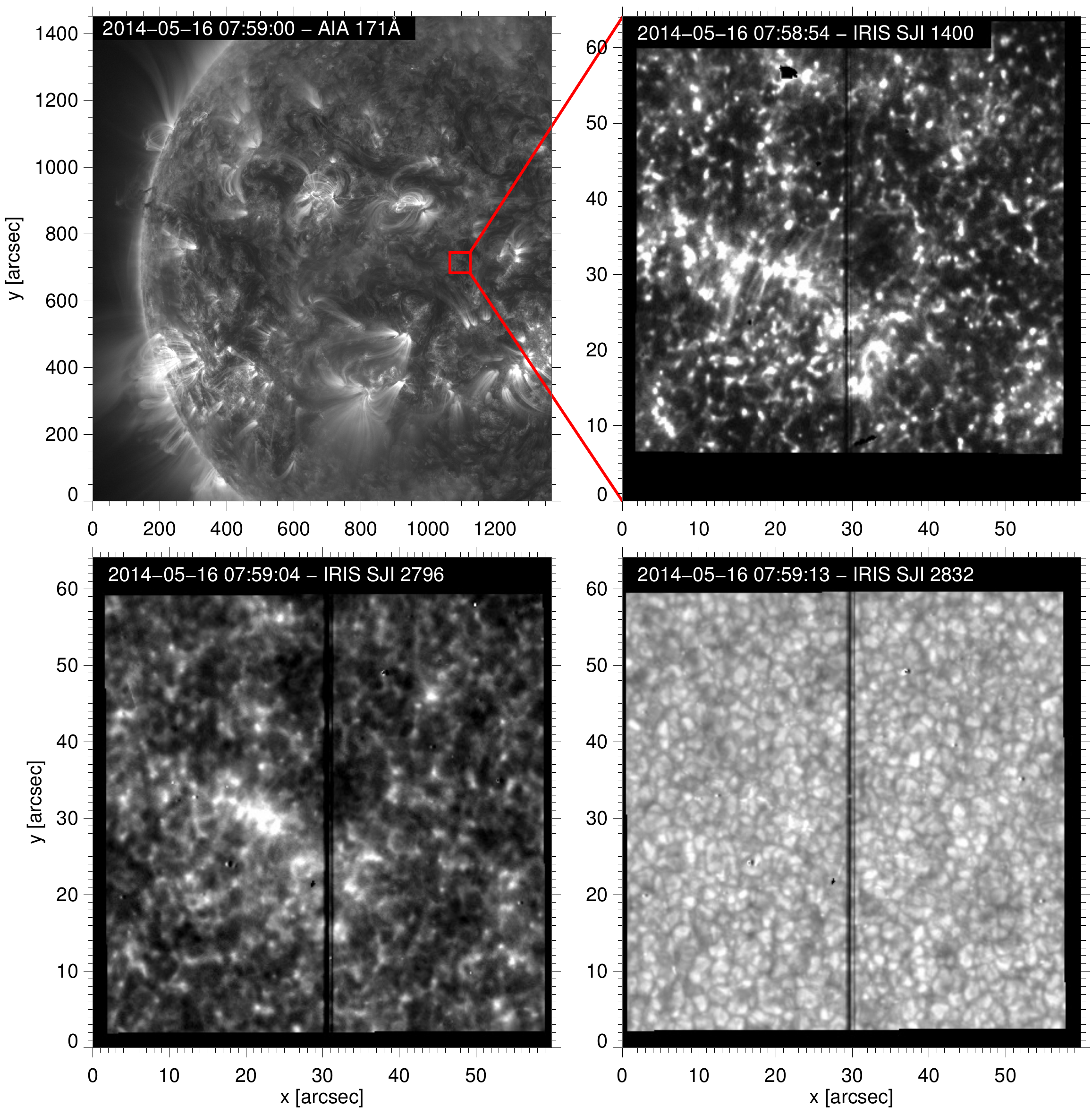}}
	\end{center}
	\vspace*{-1em}
	\caption{Upper left panel: AIA filtergram at 171\AA\ showing part of the solar disk. It is taken on 16 May 2014 at 07:59 UT. The red square outlines the observed QS region that is blown-up in the upper right and the lower panels displaying QS features visible in slit-jaw images SJI 1400, SJI 2796 and SJI 2832.}
	\label{fig1}
\end{figure*}

The observations used in this paper are presented in Sect.~\ref{sect2}. We explain how cancellations are identified and tracked in Sect.~\ref{sect3}. Sect.~\ref{sect4} provides examples of canceling events, a statistical analysis, and an estimation of the total energy released through these processes. Finally, Sect.~\ref{sect5} summarizes our findings and conclusions.

\section{Observations and data processing}
\label{sect2}

The observations employed in this work were obtained on 2014 May 16 starting at 07:23:41~UT. They consist of coordinated IRIS data sequences and SST measurements showing the evolution of QS fields at the disk center under good seeing conditions. More precisely, the target was a supergranular cell at the disk center containing small IN patches and a strong negative-polarity NE structure at the cell boundary.    

\subsection{IRIS observations}

IRIS provides spectra in three passbands, in the near ultraviolet band (NUV) from 2783 to 2834~\AA\, and in the far ultraviolet, both from 1332 to 1358~\AA\ (FUV 1), and from 1389 to 1407~\AA\ (FUV 2). This makes it possible to probe different layers of the solar atmosphere: the photosphere, the chromosphere, the upper chromosphere/lower transition region and the transition region, respectively. In addition, IRIS is capable of recording slit-jaw images (SJI) using filters centered on \ion{Mg}{2} k 2796~\AA\ (SJI 2796), the far \ion{Mg}{2} h wing at 2832~\AA\ (SJI 2832), \ion{C}{2} 1330~\AA\ (SJI 1330), and \ion{Si}{4} 1400~\AA\ (SJI 1400). The former two SJIs contain contributions from the upper photosphere (2832~\AA ) and from the upper photosphere to the upper chromosphere (2796~\AA ). For more details, we refer the reader to the first two papers in this series on the formation of IRIS diagnostics \citep{Leenaartsetal2013a, Leenaartsetal2013b}. The SJI 1400 filter is sensitive to emission from the transition region \ion{Si}{4} 1394/1403~\AA\ lines and continuum formed in the upper photosphere/lower chromosphere. Distinguishing between these two contributions can be challenging and requires analysis of multi-instrument, multi-wavelength observations \citep{MartinezSykoraetal2015}. However, whenever a given region of the solar disk is not covered by the IRIS slit (i.e., there are no available spectra), some other methods have to be used. One possible approach to separate the two contributions, based on the temporal evolution of SJI 1400 bright features, is described in Sect.~\ref{sect33}. The SJI 1330 is dominated by the \ion{C}{2} 1334/1335~\AA\ lines formed in the upper chromosphere/lower transition region and continuum formed in the upper photosphere/lower chromosphere \citep{RathoreCarlsson}.

Our observations are very sensitive, medium sparse 2-step raster data taken from 07:58:54 until 11:05:32~UT without solar rotation tracking. The cadence of the spectral observations was 18.6~s (9.3~s per raster step) while the exposure time of individual raster scans was 8~s. The slit sampled a quiet region of $0\farcs33 \times 60\arcsec$ with $1\arcsec$ as raster step size. At the same time, we took slit-jaw images (spatial pixel-size is $0\farcs16$) using the 1400~\AA\ and 2796~\AA\ filters every $\sim19$~s. Every $6^{\text{th}}$ 1400~\AA\ frame was replaced by a slit-jaw image at 2832~\AA, giving a cadence for that wavelength of 112 s. Those properties make the data ideal to study the highly dynamical IN magnetic fields. The observed QS area with its surroundings can be seen in Figure~\ref{fig1}. The upper left panel provides the context and shows the Sun as seen at 171~\AA\ with the Atmospheric Imaging Assembly \citep[AIA;][]{Lemenetal2012} onboard SDO. The other three panels show the slit-jaw images SJI 1400, SJI 2796 and SJI 2832 of the observed QS region. 

The data used here are IRIS level 2 data, meaning that dark current and flat-field corrections, and geometric and wavelength calibrations have been applied. We also performed absolute calibration of the IRIS spectra. It converts the measured intensities given in units of data number per second (DN/s) into absolute intensities expressed in erg~s$^{-1}$~cm$^{-2}$~sr$^{-1}$\AA$^{-1}$, which can be transformed into nW~m$^{-2}$~sr$^{-1}$~Hz$^{-1}$. For a detailed description of the calibration process we refer the reader to the IRIS user guide\footnote{http://iris.lmsal.com/itn26/}.

\subsection{SST observations}

The IRIS data are accompanied by coordinated observations taken at the SST with the CRisp Imaging SpectroPolarimeter \citep[CRISP;][]{Lofdahl2002, Scharmeretal2008}. CRISP is a Fabry-P\'erot filtergraph designed to obtain monochromatic images of the solar surface in selected spectral lines, from 5000 to 8600 \AA\/. It measured the polarization state of the light with a polarimeter consisting of two nematic liquid crystal variable retarders and a polarizing beam splitter located in front of the cameras.

A special observing sequence was prepared to ensure complete coverage of the lower solar atmosphere while reaching the highest polarimetric sensitivity possible. To that end, the capabilities of the SST were pushed to the limit. We performed full Stokes measurements in \ion{Fe}{1} 6173 \AA\/, \ion{Mg}{1} b$_2$ 5173 \AA\/ and \ion{Ca}{2} 8542 \AA\/, together with non-polarimetric scans through H$\alpha$ 6563 \AA\/. These lines were chosen to sample the lower and upper photosphere, and the lower and upper chromosphere, respectively.

The effective exposure times were increased with respect to the standard values to attain higher signal-to-noise ratios, which is especially important in the quiet Sun where magnetic fields are weak---both in the photosphere and in the chromosphere. We performed a \ion{Fe}{1} 6173 scan of 20 s duration (11 wavelength positions in steps of 28 m\AA\/, plus a continuum point at $+532$ m\AA\/), followed by a \ion{Mg}{1} b$_2$ scan lasting 11 s (7 wavelength positions in steps of 100 m\AA\/, plus two additional points at $\pm 50$~m\AA\/ from the line core and a continuum point at $-700$ m\AA\/), a \ion{Ca}{2} 8542 scan of 19 s duration (17 wavelength positions at steps of 100 m\AA\/ plus a continuum point at $+2.4$ \AA\/), and a short H$\alpha$ scan of only 5 s (21 spectral positions at steps of 100 m\AA\/, taking only intensity filtergrams). The observations consist of 211 such cycles with a cadence of 55 s, for a total duration of 3 hours. These data allow us to track IN magnetic elements both in the photosphere and chromosphere with unprecedented accuracy. Details of the observing sequence are summarized in Table \ref{tab1}. 

\begin{figure*}[t]
	\begin{center}
		\resizebox{1\hsize}{!}{\includegraphics[]{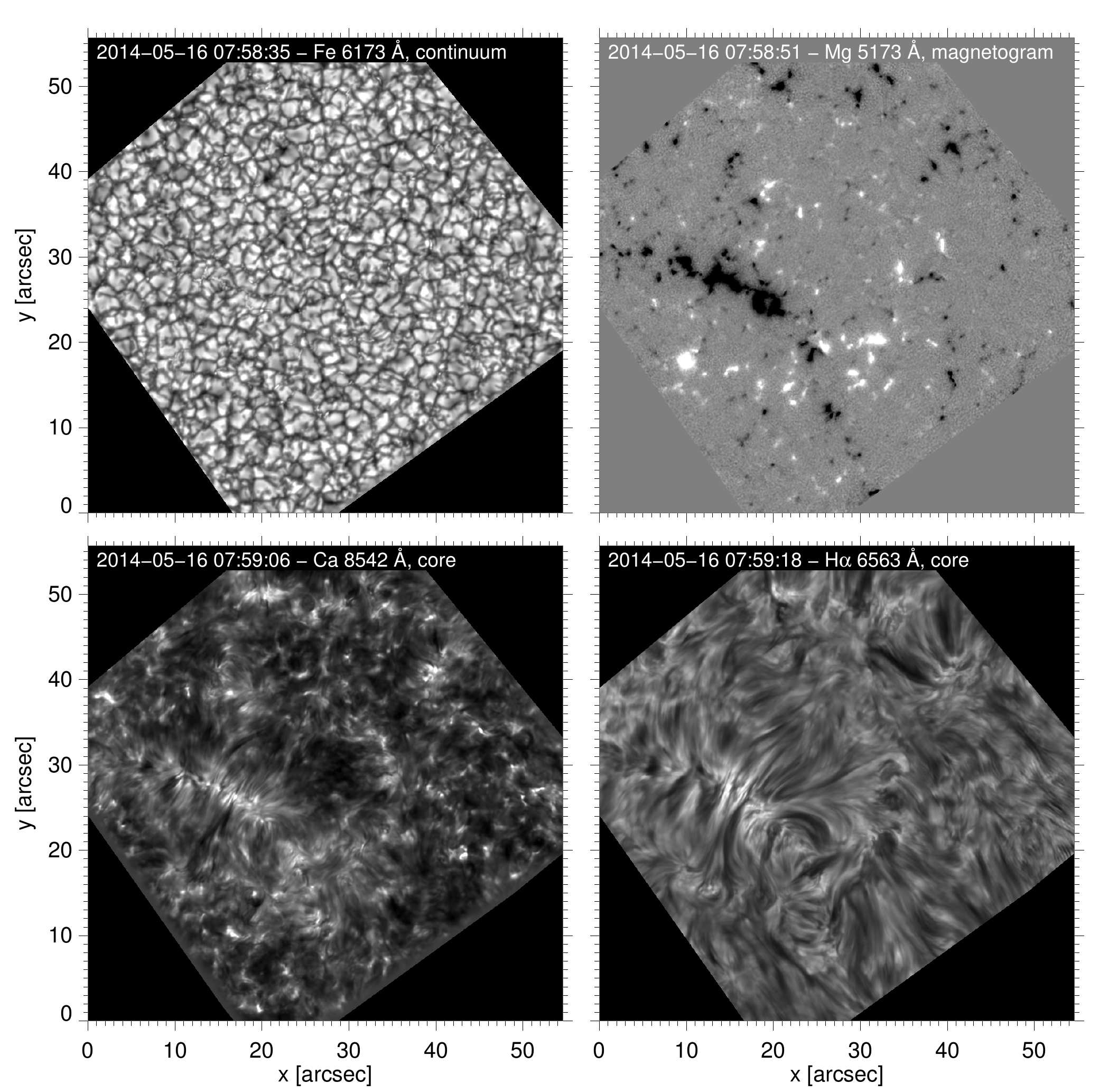}}
	\end{center}
	\vspace*{-1em}
	\caption{Examples of continuum intensity map at 6173~\AA\ (upper left), magnetogram calculated from Mg 5173~\AA\ Stokes $I$ and $V$ filtergrams at $\pm200$~m\AA\ (upper right, scaled between $\pm150$~Mx~cm$^{-2}$), and intensity maps in the core of the \ion{Ca}{2} 8542~\AA\ line (lower left), and H$\alpha$ 6563~\AA\ line (lower right). The observations were taken on May 16, 2014 at approximately 07:59~UT.}
	\label{fig2}
\end{figure*}

\begin{deluxetable*}{lcccc}
	\tablewidth{\textwidth}
	\tablecolumns{5} 
	\tablecaption{Description of the SST observations. For each observed spectral region we list their
		effective Land\'{e}-g factors (g$_{\rm eff}$), number of wavelengths ($N_{\lambda}$) and their positions with respect to the corresponding resting line centers ($\Delta\lambda$), continuum point, number of accumulations ($n$) and scan times ($t_{\text{scan}}$).\label{tab1}} 
	\tablehead{ \colhead{} & \colhead{\ion{Fe}{1} 6173 \AA\/} & \colhead{\ion{Mg}{1} b$_2$ 5173 \AA\/} & \colhead{\ion{Ca}{2} 8542 \AA\/} & \colhead{H$\alpha$ 6563 \AA\/}} \startdata
	g$_{\rm eff}$ & 2.5 & 1.75 & 1.1 & 1.05 \\
	$N_{\lambda}$ & 12 & 10 & 18 & 21 \\
	$\Delta\lambda$ (m\AA\/) & $28$ & $100$, $50$ & $100$ & $100$ \\
	Continuum point (m\AA\/) & $+532$ & $-700$ & $+2400$ & - \\
	$n$ & 14 & 9 & 9 & 8\\
	$t_{\text{scan}}$ (s) & 20 & 11 & 19 & 5\\
	\enddata
\end{deluxetable*}

\begin{figure}[t]
	\begin{center}
		\resizebox{1\hsize}{!}{\includegraphics[]{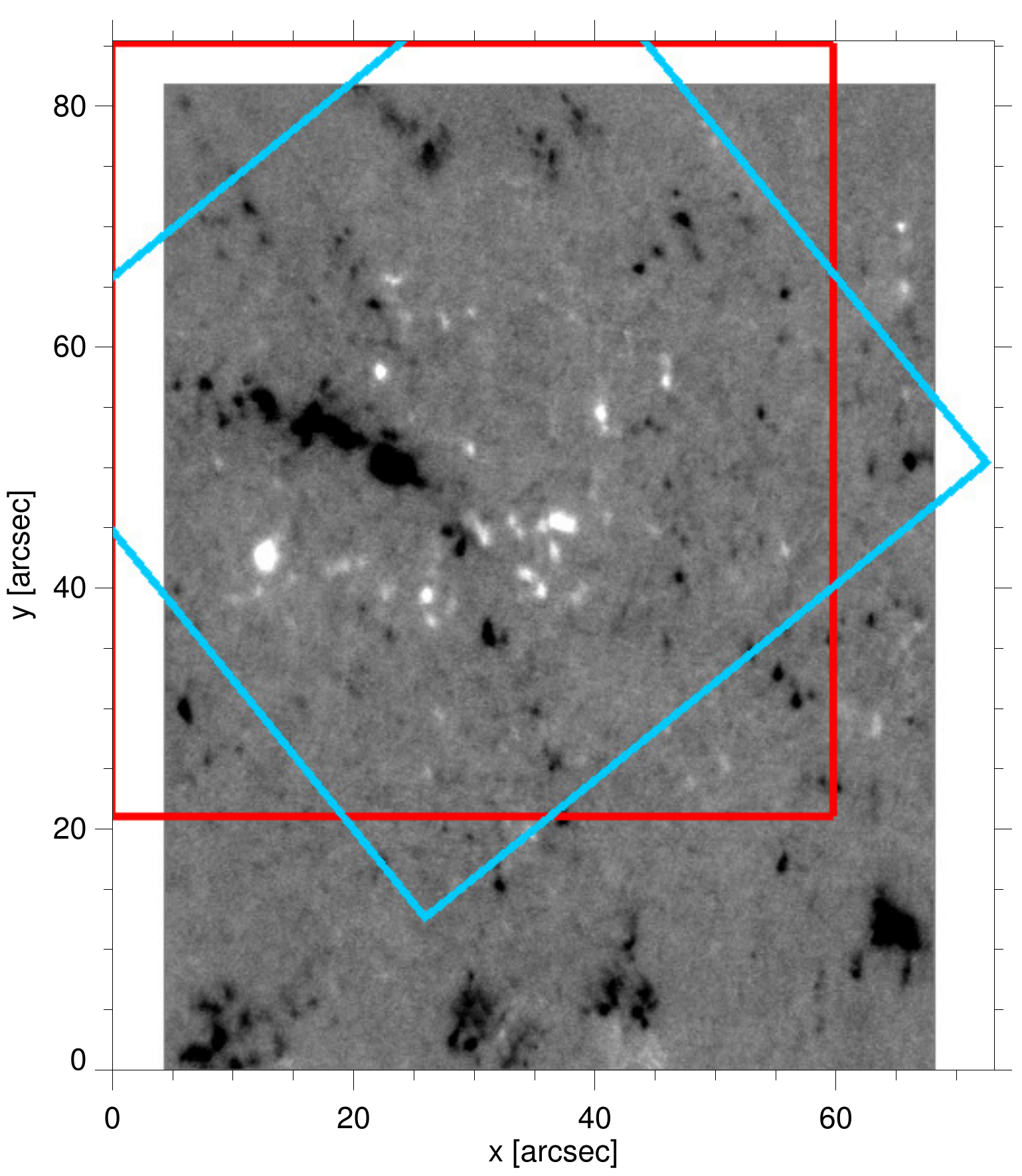}}
	\end{center}
	\vspace*{-2em}
	\caption{Maximum overlapping of the IRIS slit-jaw (red box) and the SST (blue box) FOVs at the start of the IRIS observations on 2014 May 16 at 07:58:54~UT. The background image is a circular polarization map in \ion{Mg}{1} b$_{2}$ 5173 \AA\/ obtained by the Hinode/NFI at 08:00:34~UT.}
	\label{fig15}
\end{figure}
	
Examples of the SST observations are shown in Figure \ref{fig2}. The pixel size is $0\farcs057$, sufficient to critically sample the diffraction limit of $0\farcs16$ at 6300 \AA\/. The seeing conditions were excellent for the most part, with $r_0$ values of up to 50~cm, and the SST adaptive optics system helped acquire a very stable time sequence. The observations were taken from 07:23:41 to 10:28:44 UT and monitored the same quiet Sun area. Therefore, the temporal overlap between the SST and IRIS is 2.5 hours with a continuously decreasing common field of view from roughly $50\arcsec \times 50\arcsec$ to $28\arcsec \times 50\arcsec$, as shown in Figure \ref{fig15}.

The observations were reduced using the CRISPRED pipeline \citep{delaCruzRodriguezetal2015b} and the images restored by means of the Multi-Object, Multi-Frame Blind-Deconvolution techique \citep[MOMFBD;][]{vanNoortetal2005}. Residual seeing motions were corrected employing the cross-correlation method of \cite{Henriques2012}. The pipeline includes additional corrections by \cite{vanNoortRouppevanderVoort2008}, and \cite{Shineetal1994}. To remove instrumental polarization, a telescope polarization model was used. Finally, residual crosstalk from $I$ to $Q$, $U$, and $V$ was corrected using the Stokes parameters recorded in the continuum.

Longitudinal magnetograms $\text{M}$ were computed from the observed Stokes profiles as:
\begin{equation}
\label{mag_eq}
\text{M}=\frac{1}{2}\left(
\frac{V_{\text{blue}}}{I_{\text{blue}}}-\frac{V_{\text{red}}}{I_{\text{red}}}
\right),
\label{eq1}
\end{equation}
\noindent where ``blue'' refers to the measurements in the blue wing of the line and ``red'' to those in the red wing. Thus, to first order the magnetograms are independent of LOS velocities. For example, to construct Mg 5173 magnetograms, we used filtergrams at $\pm 200$ \AA\/ from the line core. The magnetogram signal $\text{M}$ was transformed into longitudinal magnetic flux density $\phi$ using the weak-field approximation,
\begin{equation} \phi = - \frac{1}{C} \,
\frac{I(\lambda)}{{\rm d}I(\lambda)/{\rm d}\lambda}, 
\end{equation}
where $C=4.67 \times 10^{-3} \, g_{\rm eff} \lambda_0^2$, $\lambda_0$
is the central wavelength (in \AA\/) and $g_{\rm eff}$ is the effective Land\'e factor. The numerical values of $I(\lambda)$ and ${\rm d}I(\lambda)/{\rm d}\lambda$ were derived from the average quiet Sun intensity profile. To complement the longitudinal magnetograms, we also computed maps of linear polarization signal as ${\rm LP}= \sum_{i=1}^{4} \sqrt{Q(\lambda_i)^2 + U(\lambda_i)^2}/I(\lambda_i)/4$, taking into account the first four wavelengths.

For all lines, Dopplergrams were constructed as the difference of the blue and red wing intensities divided by their sum, i.e.,
\begin{equation} 
D = \frac{I_{\rm blue}-I_{\rm red}}{I_{\rm blue}+I_{\rm red}}.  
\end{equation} 
\noindent This method provides a qualitative estimate of the line-of-sight (LOS) velocity, even for the \ion{Ca}{2} 8542 line which may show complex shapes when it turns into emission.

The alignment of the various datasets was carried out as follows. First, the IRIS slit-jaw images were compensated for solar rotation and scaled up to match the SST pixel size. We then aligned the IRIS and SST observations using prominent NE features and bright points in SST \ion{Fe}{1} 6173 continuum intensity and IRIS SJI 2832 images. Since both channels practically show the same photospheric structures, the accuracy of the alignment is on the order of the IRIS pixel size. The other SST and IRIS channels were aligned to these two channels. Finally, the SST data and the IRIS SJI 1400 and SJI 2832 images were interpolated in time applying the nearest neighbor method of interpolation to match the cadence of the SJI 2796 images (19~s).

We made extensive use of CRISPEX \citep{VissersRouppevanderVoort2012} to visualize, analyze, and interpret the data. CRISPEX is part of the IDL SolarSoft package\footnote{SolarSoft is a set of integrated software libraries, data bases, and system utilities which provide a common programming and data analysis environment for solar physics and can be accessed at http://www.lmsal.com/solarsoft/}.
    
\section{Method}
\label{sect3}

To understand how cancellations of small-scale IN fields are coupled to the dynamics and energetics of the upper solar atmosphere, we have detected and tracked all the individual magnetic elements visible in the photospheric data. From these patches, we identified and selected only those involved in cancellation processes and analyzed how they affect the chromosphere and corona. To that aim, we also retrieved various atmospheric parameters employing an inversion code, that we later used to estimate the energy budget of the canceling events. We described below each of these steps in detail. 

\subsection{Identification and Tracking of Magnetic Features in SST magnetograms}
\label{sect31}

To determine the history of individual flux features in the solar photosphere, the most suitable SST magnetograms are those constructed from Stokes $I$ and $V$ filtergrams taken at $\pm200$~m\AA\ from the core of the Mg 5173~\AA\ line. Individual elements were automatically identified and tracked using the YAFTA code \citep{WelschLongcope2003}, applying the clumping method. The adopted magnetogram signal threshold is $3\sigma$, where $\sigma$ (8~Mx~cm$^{-2}$) is the standard deviation calculated in a region without clear solar signals. We consider as real magnetic concentrations only those with a minimum size of 16 pixels. Magnetic features that appear and disappear in-situ have to be visible in at least two frames. We corrected the YAFTA tracking results using the code developed by \cite{GosicPhD} which resolves the errors made by YAFTA during the tracking process. This code allows us to properly interpret interactions between flux elements and intrinsic changes in the element signals.

\subsubsection{Separation of IN from NE magnetic elements}

\begin{figure}[t]
	\begin{center}
		\resizebox{1\hsize}{!}{\includegraphics[]{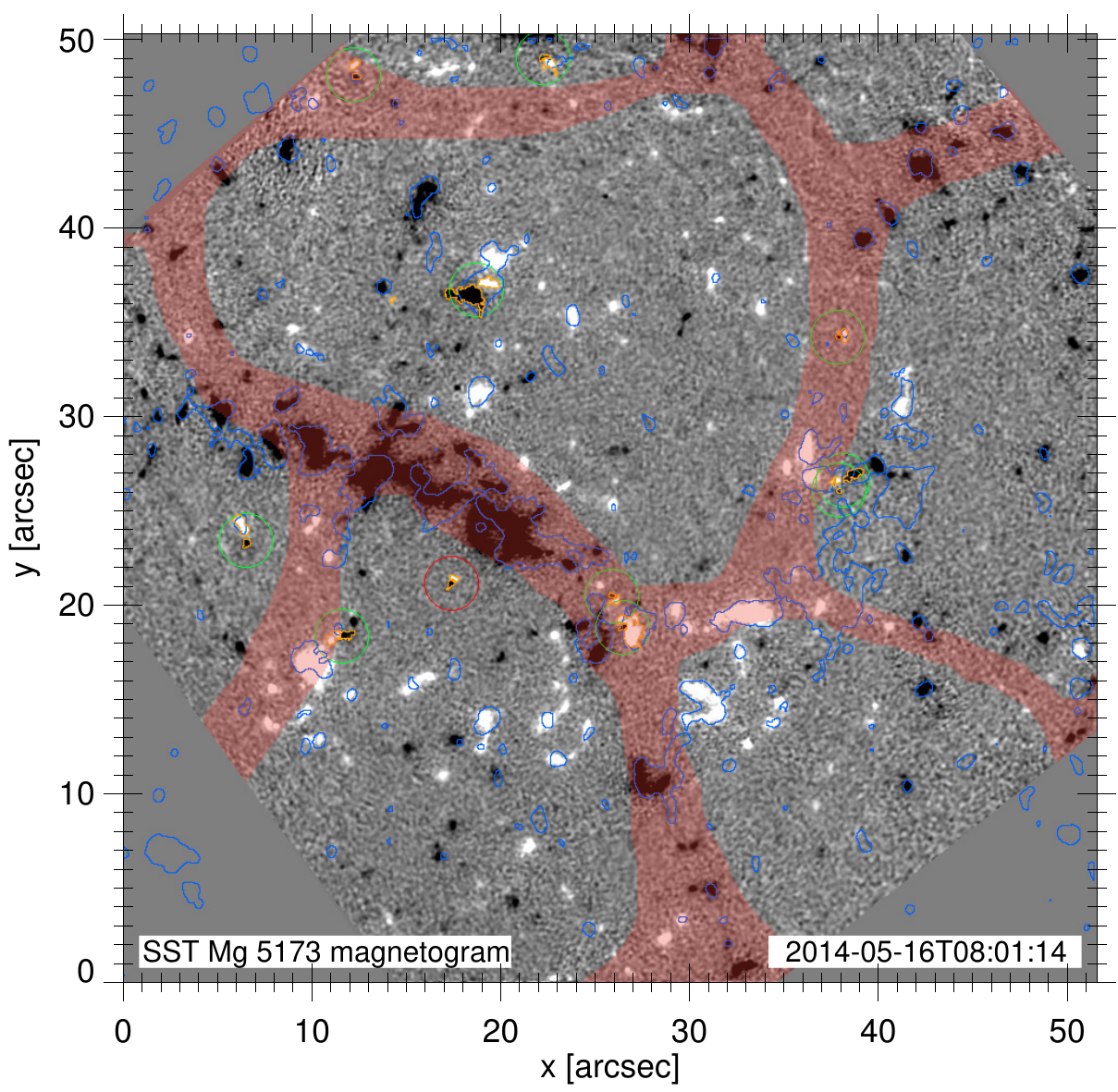}}
	\end{center}
	\vspace*{-1em}
	\caption{NE and IN regions as determined by the LCT technique applied to SST Fe 6173 Dopplergram maps. The red shaded area coincides with the NE while the non-shaded zones represent the IN, i.e., the interior of supergranular cells.}
	\label{fig14}
\end{figure}

In order to distinguish between IN and NE flux patches we have applied the same method as described in \cite{Gosicetal2014}. In this way, the IN regions are considered to coincide with the interior of supergranular cells while the space beyond the IN are the NE regions. Supergranular cells are determined by applying the local correlation tracking (LCT) technique \citep{November} on SST Fe 6173 Dopplergram maps, considering the whole time sequence of 3 hr. The LCT algorithm calculated horizontal velocities employing a Gaussian tracking window of FWHM of 8.5 arcsec, which is large enough to suppress small convective patterns such are granular flows while preserving the large-scale supergranular motions. Figure \ref{fig14} shows separated NE and IN regions, with the red shaded area representing the NE.

Magnetic elements that were in the interior of supergranular cells at the beginning, and those that appeared inside supergranular cells during the observational period, were taken to be IN elements. All other magnetic patches were considered to be NE structures.

\subsection{Cancellation events}
\label{sect32}

Cancellation of opposite-polarity flux patches in the photospheric magnetograms lead to slow fading of the patches involved in the interaction. Therefore, YAFTA does not see the difference between cancellation and in-situ disappearance. For this reason, to detect cancellations, we follow the same strategy as described in detail in \cite{Gosicetal2014}. To summarize, we use the YAFTA output and search for all IN elements that disappear in situ in the current frame. The borders of these elements are dilated by four pixels and those that overlap opposite polarity patches are identified as canceling features. Each detected pair of canceling elements form a canceling event to which we assign a unique label. Using these labels we can study the history of individual canceling events.  

In addition, magnetic elements may undergo partial cancellations as well, which also need to be identified. The reason is that, although flux patches do not disappear through these processes, it is possible that reconnection occurs and releases magnetic energy. Partial cancellations are detected similarly to the previous case. We first expand borders of each IN element that does not disappear in a given frame by four pixels. If a given element overlaps with an opposite polarity patch, we inspect whether they are in contact because they form a newly emerging magnetic loop. The latter is done by checking when and how flux patches appear and determine in what direction they are moving. Those patches that appear in situ, that have been visible in the magnetograms in less than 5 frames and then move in opposite directions are considered as bipolar structures and discarded from the further study. Otherwise, they are taken as flux features involved in a partial cancellation.

\subsection{Association of IN cancellations with bright grains in SJI 1400}
\label{sect33}

Cancellation of IN fields is believed to be associated with transient brightenings in the chromosphere and/or transition region. In such cases, we expect to detect signal in SJI 1400 above cancellation sites, which may be a signature of reconnection processes that release energy and/or hot plasma. 

However, special care must be taken with the detection of SJI 1400 bright structures, because not all of them originate from reconnection of magnetic field lines. Sometimes, bright grains come from the continuum intensity in the SJI 1400~\AA\ filters. They are normally formed in the chromosphere due to upward propagating acoustic waves in non-magnetic environment and sometimes they can even reach the transition region \citep{MartinezSykoraetal2015}. These shock-related grains are highly transient features. They move fast and have lifetimes of roughly two minutes. Hence, the grains can be removed from the IRIS 1400 SJI images by applying a subsonic filter \citep{Titleetal1989, Strausetal1992} with 5~km~s$^{-1}$ as the cutoff. This makes association of IN cancellations with brightenings in the transition region more reliable.  

To check which IN cancellations are interrelated with SJI 1400 brightenings, we first identify these brightenings using YAFTA and the downhill method, considering all the pixels in SJI 1400 with signal above a threshold level of 60 counts per pixel. We decided to use these settings after careful visual inspection. We have concluded that with such a setup we do not miss features that we believe are real, i.e., they are persistent and clearly distinguishable from the background noise. We will discuss in Section \ref{sect4} how different parameters for the detection of SJI 1400 features affect the results. After the identification is done, we find the intersections of the canceling magnetic elements detected in the magnetograms by dilating their borders by four pixels. This is done for each canceling event and in each frame. When an intersection overlaps with any of the detected patches in SJI 1400, we mark the corresponding canceling event as the one associated with brightenings in the chromosphere/transition region.

\begin{figure*}[t]
	\begin{center}
		\resizebox{1\hsize}{!}{\includegraphics[]{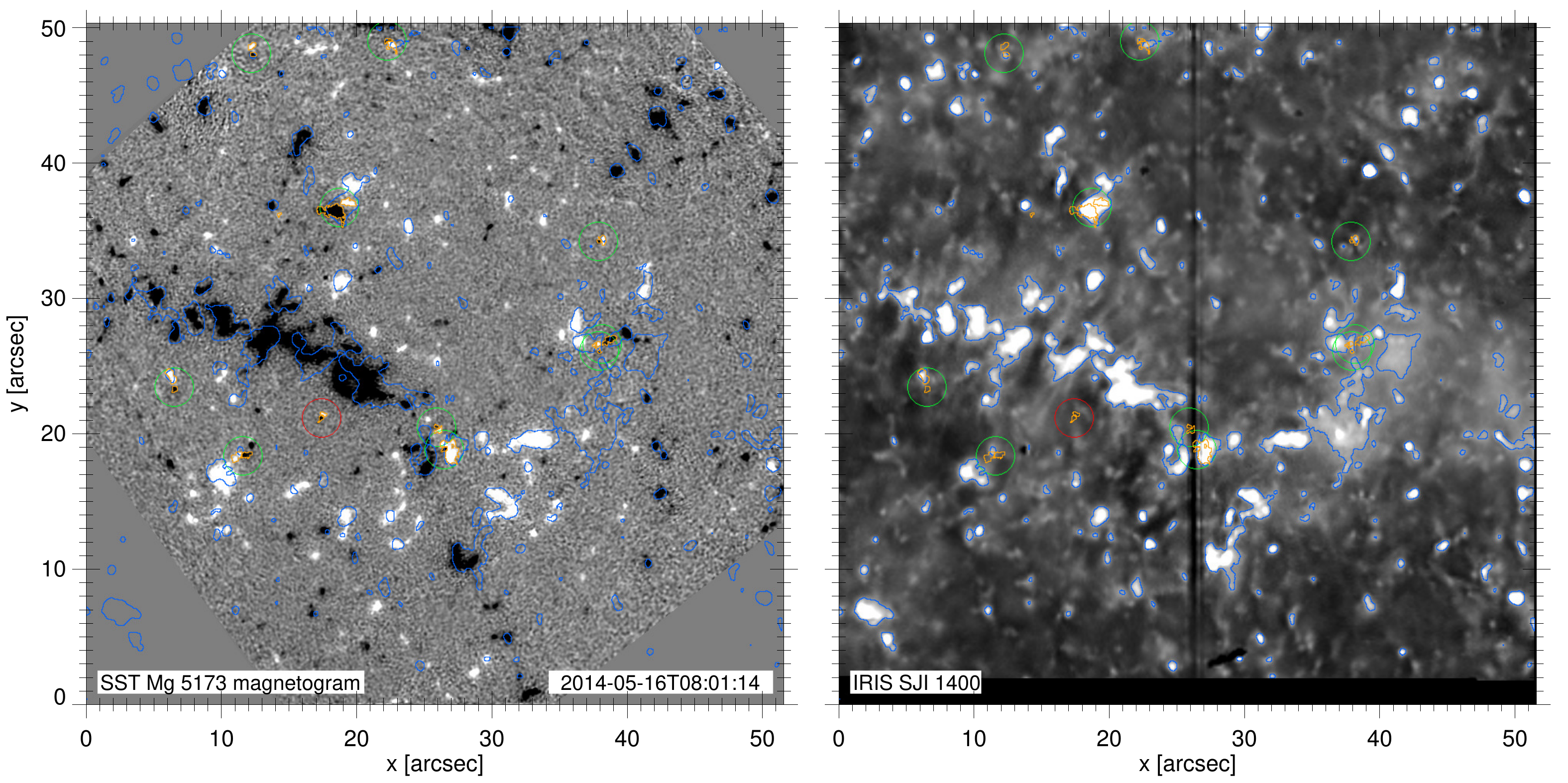}}
	\end{center}
	\vspace*{-1em}
	\caption{Snapshot from an animation of Mg 5173~\AA\ magnetogram showing IN flux patches that
		are completely or partially canceling in the upper photosphere (left panel, orange contours). Above some of the canceling events (enclosed with green circles), SJI 1400 bright features (marked with blue contours) are clearly visible. Those cancellations that do not have an associated bright grain in SJI 1400 are marked with red circles. The corresponding SJI 1400 image is shown in the right panel.\newline
		{\em An animation of this figure is available in the online journal.}}
	\label{fig3}
\end{figure*}

\begin{figure*}[!t]
	\begin{center}
		\resizebox{1\hsize}{!}{\includegraphics[]{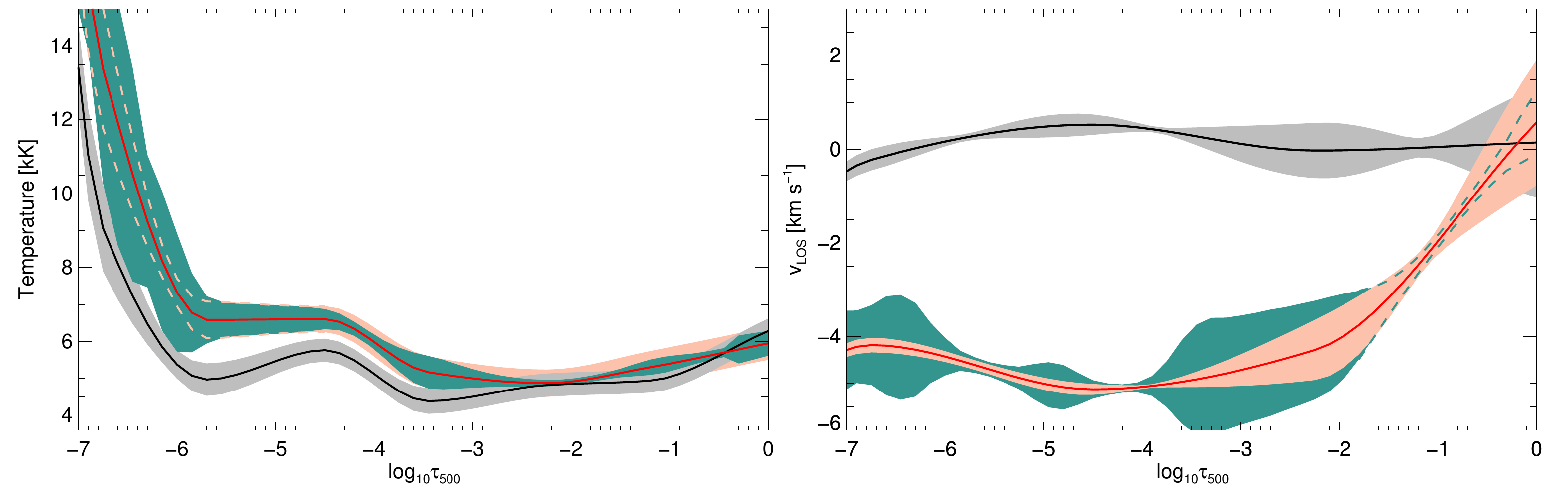}}
	\end{center}
	\vspace*{-1em}
	\caption{Uncertainties inferred from the inversions of the \ion{Mg}{2} h \& k and UV triplet lines. The solid black (a QS pixel) and red (pixel within a canceling region) curves show the temperature (left panel) and LOS velocity (right panel) distributions. The gray and red shaded areas cover their respective uncertainty regions. In addition to this, the green shaded areas around the curves representing the canceling pixel show uncertainties derived from the corresponding response functions.}
	\label{fig16}
\end{figure*}

The quality of the final tracking and association of canceling events with the SJI 1400 signal can be evaluated from the animation of Figure \ref{fig3}. It shows cancellations detected in the Mg 5173~\AA\ magnetograms (left panel). Flux patches undergoing cancellations have orange contours during the process, otherwise their borders are not drawn. If canceling events have associated signal in SJI 1400 they are outlined with green circles, and with red if not. Blue contours enclose SJI 1400 pixels where the signal is above 60 counts per pixel (as shown in the right panel).   

\subsection{Inversions of IRIS data}
\label{sect34}

IRIS provides unprecedented time sequences of \ion{Mg}{2} spectra at high spatial, spectral, and temporal resolution. These spectra represent the most promising IRIS chromospheric diagnostics, permitting us to infer thermodynamical properties of the solar atmosphere \citep{Leenaartsetal2013b, Pereiraetal2015}. We used STiC \citep{delaCruzRodriguezetal2016} to perform chromospheric data inversions. STiC is a new inversion code based on the RH synthesis code \citep{Uitenbroek2001} and it allows to model spectral lines in non-LTE including the effect of partial redistribution in angle and frequency of scattered photons.

We apply the STiC code to our IRIS \ion{Mg}{2} spectra in order to retrieve temperature, line-of-sight velocities ($v_{\rm los}$), microturbulence, gas density, gas pressure, and electron density as functions of optical/geometrical depth at IN cancellation sites. These physical parameters will be used to examine the chromospheric response to IN cancellations. To this purpose, we selected every third pixel along the slit at each third time step, and run the inversions in intensity-only mode. This way we created a relatively coarse map of atmospheric parameters of the QS scanned by the slit. The inversions were carried out in two cycles, ending with 9 nodes in temperature, 5 nodes in $v_{\rm los}$ and 3 nodes in microturbulence. Through trial-and-error we found that these are the minimum numbers of nodes that provide reasonably good fits to the IRIS QS \ion{Mg}{2} h \& k and UV triplet lines. The gas pressure at the upper boundary of the model atmosphere is also considered as a free parameter. As initial atmosphere we used the FALC model \citep{Fontenlaetal1993}. The initial model contains a given stratification of temperature, $v_{\rm los}$, microturbulence, and the gas pressure as a function of optical depth. To get sufficient emission in the line cores, we increased the FALC gas pressure by an order of magnitude. The second cycle is initialized employing the model atmosphere obtained in the first cycle. To derive a more reliable photospheric $v_{\rm los}$, we included the photospheric \ion{Ni}{1} 2814.350~\AA\ line following the strategy from \cite{delaCruzRodriguezetal2016}, which is modelled in LTE.

The uncertainties in temperature and LOS velocity that can be expected from the inversions are shown as the gray and red shaded areas in Figure \ref{fig16}. They are computed as the standard deviations of the corresponding atmospheric parameters obtained through inversions of the selected IRIS lines in two pixels using one thousand randomly perturbed initial model atmospheres. One pixel is located within a non-magnetic QS region (black lines) while the other one represents a canceling region (red lines). 

Another way to calculate the uncertainties of the model is given by Eq. (42) in \cite{DelToroIniestaRuizCobo}:
	
	\begin{equation}
	\sigma^{2}_{p}\simeq\frac{2}{nm+r}\frac{\sum\limits_{s=0}^3\sum\limits_{i=1}^q[I_{s}^{syn}(\lambda_{i})-I_{s}^{obs}(\lambda_{i})]^{2}w_{s,i}^{2}}{\sum\limits_{s=0}^3\sum\limits_{i=1}^qR_{p,s}^{2}(\lambda_{i})w_{s,i}^{2}},
	\end{equation}	
	
\noindent where, $n$ is the number of optical depth points, $m$ is the number of physical quantities varying with depth while $r$ represents those that are constant with height. $s$ and $i$ scan over the four Stokes parameters and $q$ wavelength samples, respectively. $I_{s}^{syn}$ and $I_{s}^{obs}$ are the synthetic and the observed intensity profiles. $R_{p,s}$ are the response functions of a given Stokes parameter to perturbations of the atmospheric quantity $p$, which runs from $1$ to $nm+r$. Finally, $\omega_{s,i}$ stands for the weights of the Stokes parameters. The uncertainties are shown in Figure \ref{fig16} as the green shaded areas around the curves representing the canceling pixel, and for clarity, they are omitted for the QS pixel. As can be seen, they are of the same order of magnitude as the uncertainties derived from the Monte Carlo approach.

\begin{figure*}[t]
	\begin{center}
		\resizebox{1\hsize}{!}{\includegraphics[]{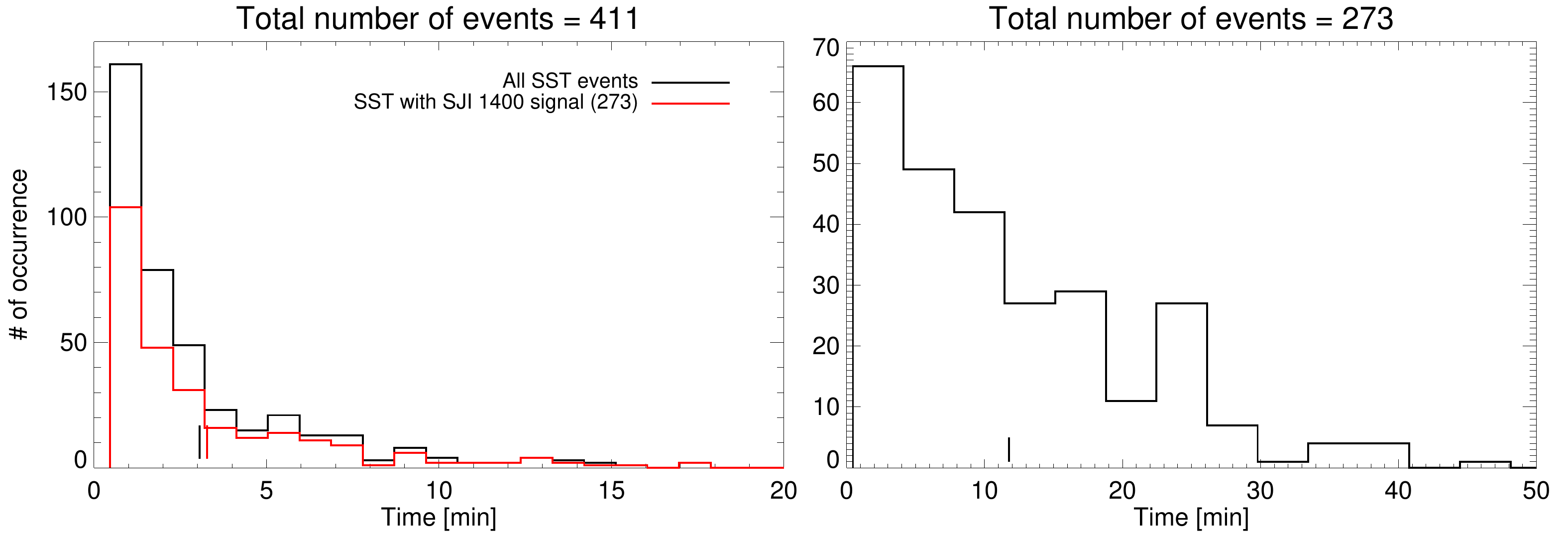}}
	\end{center}
	\vspace*{-1em}
	\caption{\textit{Left panel}: Lifetimes of all canceling events detected in the Mg 5173~\AA\ magnetograms (black line) and those that have associated SJI 1400 signal (red line). The bin sizes are $55$~s. \textit{Right panel}: Lifetimes of SJI 1400 bright grains assigned to cancellations in Mg 5173~\AA\ magnetograms (the bin size is $220$~s). The black and red vertical lines mark the mean values of the corresponding lifetime distributions.}
	\label{fig9}
\end{figure*}

\begin{figure}[t]
	\begin{center}
		\resizebox{1\hsize}{!}{\includegraphics[]{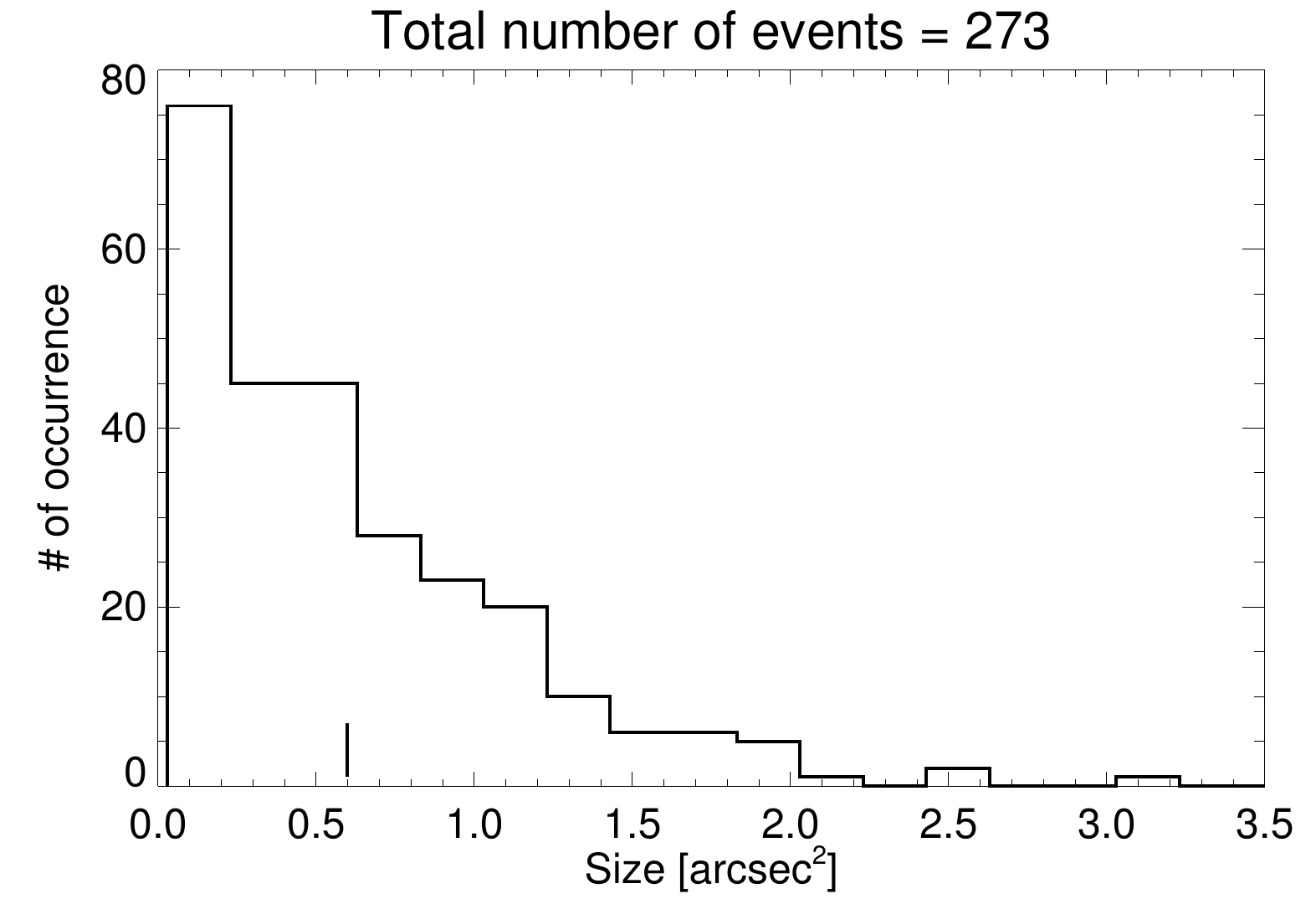}}
	\end{center}
	\vspace*{-1em}
	\caption{Size distribution of IRIS SJI 1400 bright grains. The bin sizes used are $0.2$~arcsec$^{2}$. The black vertical line indicates the mean value of the distribution.}
	\label{fig10}
\end{figure}

\section{Results}
\label{sect4}

Using the corrected YAFTA output we identified all total and partial cancellations of flux patches in Mg 5173~\AA\ magnetograms. In total, we detected $411$ canceling events ($592$ magnetic elements) of which $32$ include only NE patches. However, since NE cancellations play only a minor role in our observations and for the sake of simplicity, we will consider all canceling events as IN cancellations further in our analyses. This accounts for $0.1$ elements per arcsec$^{2}$ and per hour. Many cancellations show the standard evolutionary pattern. Two opposite polarity patches approach each other and start decreasing in size and flux. If the cancellation is total, they completely disappear from the magnetograms. Otherwise, either one or both elements survive the interaction. The interacting magnetic features can also be seen in intensity filtergrams as two tiny bright structures located in intergranular lanes that move toward each other. This is accompanied by increased brightening in IRIS slit-jaw and the Ca~II~8542 \AA\ images. Some cancellations undergo much more complex evolution, which includes surface processes such as fragmentations and mergings of flux patches.

If we discard $51$ canceling events that are not covered by IRIS due to decreasing overlapping FOV between the SST and IRIS observations, $76\%$ of the detected canceling events are cospatial with SJI 1400 signals. The signal-weighted centers of the bright grains are on average by $0\farcs5$ away from the corresponding cancellation site centers (defined as the center of intersections between interacting opposite-polarity flux patches). This small dislocation is likely the result of the alignment uncertainty. As mentioned before, SJI 1400 bright features are detected taking into account all the pixels above the threshold level of 60 counts per pixel. When the threshold and minimum size of SJI 1400 features are set to 70 counts and 4 pixels, respectively, $70\%$ of the canceling events overlap with SJI 1400 structures. This ratio increases to $87\%$ if all pixels above 40 counts are considered. Too restrictive selection criteria imply that many real signals will remain undetected. On the other hand, too low thresholds yield numerous identifications of spurious features. Since the ratio between these two limiting scenarios does not change by more than $10\%$, we are confident that the SJI 1400 signal we detect above the canceling regions is real. In $\sim80\%$ of cases, this signal starts to form, on average, $\sim4$ minutes before we detect cancellations in Mg 5173~\AA\ magnetograms. 

The events without associated SJI 1400 features may imply several possibilities: they are the result of our selection criteria, SJI 1400 images do not have sufficiently high sensitivity, or canceling flux features are simply too weak to affect the chromosphere and layers above. 

Hereafter, we characterize the following parameters of the detected cancellations: lifetime, size and atmospheric physical parameters (from the inversions). The latter are used to estimate the total energy released through the QS cancellations. A general description of two canceling events is provided after this characterization. We have proceeded with a detailed analysis of the second event for which we also have IRIS spectra.

\subsection{Lifetime and size}

The lifetime distributions of the detected canceling events are displayed in Figure \ref{fig9} (left panel). In our magnetogram sequences, an event is considered to live from the frame in which flux patches start to interact until they get separated or one of them disappears. The lifetimes of cancellations range from 1~min up to 22~min. The lower limit is set by the cadence of the SST observations. Both distributions, for all canceling events (black line) and for those with associated SJI 1400 bright features (red line), have practically the same mean lifetime of about $3$ minutes. 

However, SJI 1400 bright structures above canceling events live $11.8$ minutes on average (right panel in Figure \ref{fig9}), i.e., much longer than cancellations seen in Mg 5173~\AA\ magnetograms. We note that the downhill method \citep{WelschLongcope2003}, used to identify SJI 1400 bright grains, detects local maxima and classifies each of them as a distinct patch. This approach is suitable whenever it is necessary to separate individual patches from large complex structures. However, this means that the lifetimes of SJI 1400 bright grains associated with canceling events are determined by the downhill method itself, keeping in mind its limitations \citep{DeForestetal2007}.

The size distribution of the detected SJI 1400 bright grains is shown in Figure~\ref{fig10}. The lower and upper limits of the distribution are $0.03$~arcsec$^{2}$ and $3.4$~arcsec$^{2}$, respectively. The mean size of bright grains is about $0.6$~arcsec$^{2}$. Interestingly, this is similar to the mean size of IN flux concentrations measured in Hinode/NFI photospheric magnetograms \citep{GosicPhD}. 

\subsection{Examples of canceling events}

In this section we will describe in detail the evolution of two canceling events. They show how opposite-polarity IN magnetic fields interact, affecting the upper solar atmosphere up to the transition region.

\begin{figure*}[t]
	\begin{center}
		\resizebox{1\hsize}{!}{\includegraphics[]{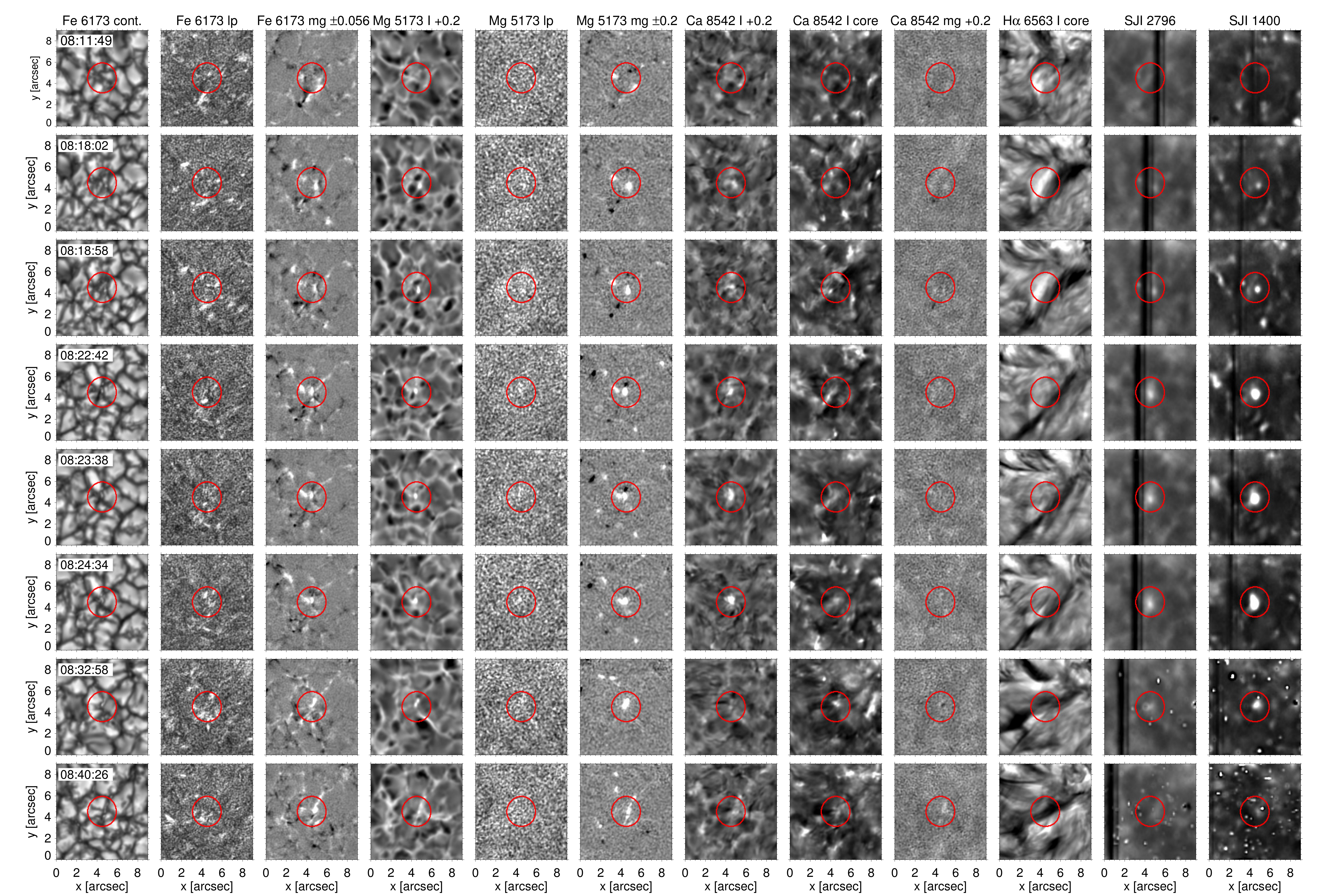}}
	\end{center}
	\vspace*{-1em}
	\caption{Temporal sequences of several SST and IRIS observables during the flux cancellation phase. From left to right: continuum intensity, linear polarization and magnetograms in \ion{Fe}{1} 6173~\AA, intensity images in the red wing of the Mg 5173~\AA\ line and the corresponding linear polarization maps and magnetograms, filtergrams in \ion{Ca}{2} 8542~\AA\ (the red wing and core),  \ion{Ca}{2} magnetograms, and the last three rows show the core in H$\alpha$ 6563~\AA, IRIS SJI 2796 and 1400 filtergrams. The red circles outline two canceling IN flux concentrations located around $(x, y)=(5\arcsec, 5\arcsec)$. Starting from 08:18:02~UT, the filtergrams showing the chromosphere and transition region reveal brightening features at the observed location. IN flux concentrations in the lower-lying photosphere (as seen in the photospheric magnetograms) are seen to be canceling. \newline
		{\em An animation of this figure is available in the online journal.}}
	\label{fig4}
\end{figure*}

\begin{figure*}[t]
	\begin{center}
		\resizebox{0.5\hsize}{!}{\includegraphics[]{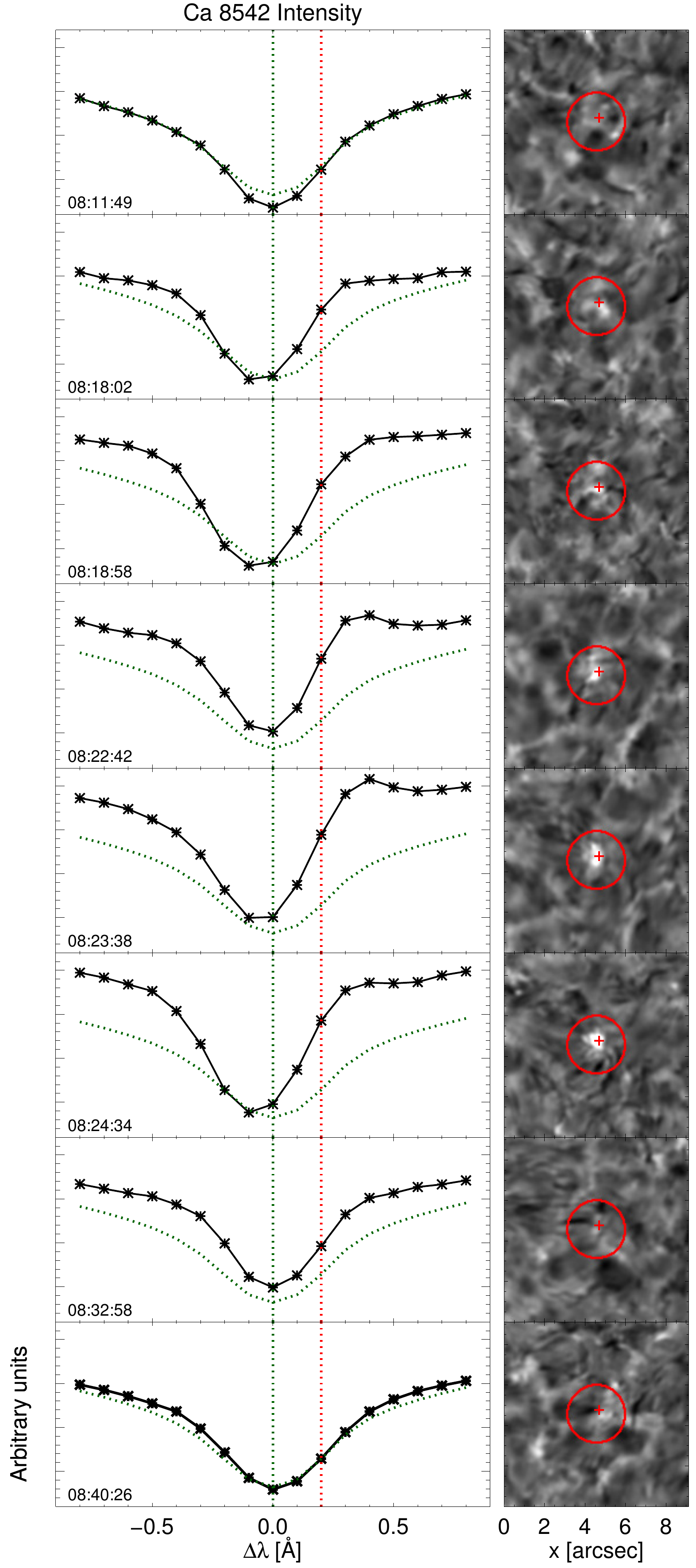}}
	\end{center}
	\vspace*{-1em}
	\caption{Temporal evolution of Stokes $I$ profiles in the \ion{Ca}{2} $8542$~\AA\ line (black lines) formed at the center of the canceling event presented in Figure \ref{fig4}. The spectra is observed at the location marked with the red cross in the filtergrams shown on the right. The dotted green curve represents the average QS profile in the accessible SST FOV. Time increases from top to bottom, as indicated in the lower left corner of each panel. The vertical dotted green and red lines in the spectra panels represent the rest wavelength of the line center (i.e., 8542 \AA) and the one at which the intensity maps in the right column are taken ($8542+0.2$~\AA), respectively.}
	\label{fig5}
\end{figure*}

\subsubsection{Example \#1}

Example 1 portrays a simple canceling scenario of two small IN flux concentrations. The time evolution for this example is displayed in Figure \ref{fig4} and the accompanying movie, where the flux canceling region is enclosed by a red contour. The figure shows twelve observables, from left to right: continuum intensity at Fe $6173$~\AA\, linear polarization and lower photospheric magnetograms taken in the same line ($\pm56$~m\AA), wing intensity (+$200$~m\AA), linear polarization and upper photospheric magnetograms ($\pm200$~m\AA) recorded in the Mg $5173$~\AA\ line, and intensity maps in the wing ($+200$~m\AA) and core of \ion{Ca}{2} $8542$ where brightenings are more pronounced. The final four rows are \ion{Ca}{2} magnetograms, intensity filtergrams in the core of H$\alpha$~$6563$~\AA\ and IRIS 2796 and 1400 slit jaw images. The associated movie displays the same observables. These temporal sequences clearly show cancellation of IN fluxes and the chromospheric response to it. 

The first sign of the chromospheric activity above the canceling region appears at 8:11:49~UT in the red wing of \ion{Ca}{2} $8542$~\AA. We see a small, faint brightening at $(x, y)=(5\farcs5, 3\farcs9)$, probably triggered by cancellation of very small negative flux patch with stronger positive flux concentration visible in Fe magnetograms. None of the other filtergrams show any activity at that location nor do Mg magnetograms show interaction of flux patches. Three minutes later a small bright structure appears in IRIS SJI 1400, coinciding with the positive magnetic element at $(x, y)=(5\farcs5, 3\farcs9)$. It becomes stronger and larger very fast while the signal at \ion{Ca}{2} $8542+0.2$~\AA\ follows a similar evolutionary pattern.

In the mean time, two negative flux patches visible in Mg $5173\pm0.2$~\AA\ magnetograms at $(x, y)=(5\farcs7, 5\farcs5)$ are being dragged by granular motions within intergranular lanes and coalesce into a larger, stronger individual flux structure at 8:20:50~UT. The newly created flux concentration migrates toward the positive element $(4\farcs5, 4\farcs5)$. The evolution of their corresponding footpoints can also be seen in the first and fourth rows in Figure \ref{fig4} as small-scale bright points (BPs) between granular cells. These BPs can better be discerned in the intensity maps in the red wing ($+200$~m\AA) of the Mg 5173~\AA\ line, especially the one corresponding to the positive polarity element. They also coincide with strong local downflows visible in Fe and Mg Dopplergrams. Simultaneously, the first hint of a somewhat permanent brightening shows up in the core of the \ion{Ca}{2} line. 

The cancellation in both Fe $6173\pm0.056$~\AA\ and Mg $5173\pm0.2$~\AA\ magnetograms starts to be visible from approximately 08:21:46~UT. By that time, intensity enhancements are easily discernible in \ion{Ca}{2} $8542+0.2$~\AA\ and IRIS slit jaw images. They attain their maximum in size and brightness from 08:23:38~UT to 08:25:12~UT when an apparent bright structure is visible in both \ion{Ca}{2} filtergrams which can be interpreted as a signature of upflowing plasma (see below). During this period, there is a weak LP signal in the blue wing of the Mg $5173$~\AA\ line. Since the signal is too weak it can easily be the result of intensity to Stokes $Q$ and $U$ cross-talk or noise fluctuations.  

The negative flux patch disappears completely from the Mg magnetograms at 08:25:30~UT. The positive patch, however, survives the interaction and continues its evolution. In Fe magnetograms we can still follow the flux cancellation at the photospheric level. The disappearance of the negative flux structure is followed by strong LP located at $(x, y)=(\sim$$4\farcs5, \sim$$5\farcs5)$, between positive and negative patches. It is possible that the detected LP feature is a signature of a newly created descending loop. In the mean time, filtergrams in the core of the H$\alpha$ 6563~\AA\ line reveal the formation of a dark cloud at 8:24:34~UT, $(x, y)=(\sim$$5\farcs7, \sim$$5\farcs5)$. The cloud transform into elongated dark structure and can be observed until 08:35:09~UT when it merges with another dark fibril coming from its right side. Bright structures in \ion{Ca}{2}, SJI 2796 and SJI 1400 filtergrams can be observed in several more time steps. They first disappear from the \ion{Ca}{2} $8542+0.2$~\AA\ at 8:30:10~UT, and from IRIS slit jaw images about 3 minutes later. Interestingly, the movie shows that transient brightenings in the core of the \ion{Ca}{2} $8542$~\AA\ line, above the canceling event, can be further observed until 08:41:03~UT. 

Figure \ref{fig5} shows the temporal evolution of Stokes $I$ profiles in the \ion{Ca}{2} $8542$~\AA\ line. The black solid curves in the left panels represent \ion{Ca}{2} intensity spectra obtained at the position marked with the red cross in the filtergrams in the right column. The dotted green curves show the average QS profile calculated in the available SST FOV, avoiding the strongest, negative network structure. Time is shown in the lower left corners and increases from top to bottom. The vertical dotted green lines in the left column mark the wavelength of the resting line center. The \ion{Ca}{2} $8542+0.2$~\AA\ wavelength is shown with a red dotted vertical line and the corresponding filtergrams are displayed in the right column.

\begin{figure*}[t]
	\begin{center}
		\resizebox{1\hsize}{!}{\includegraphics[]{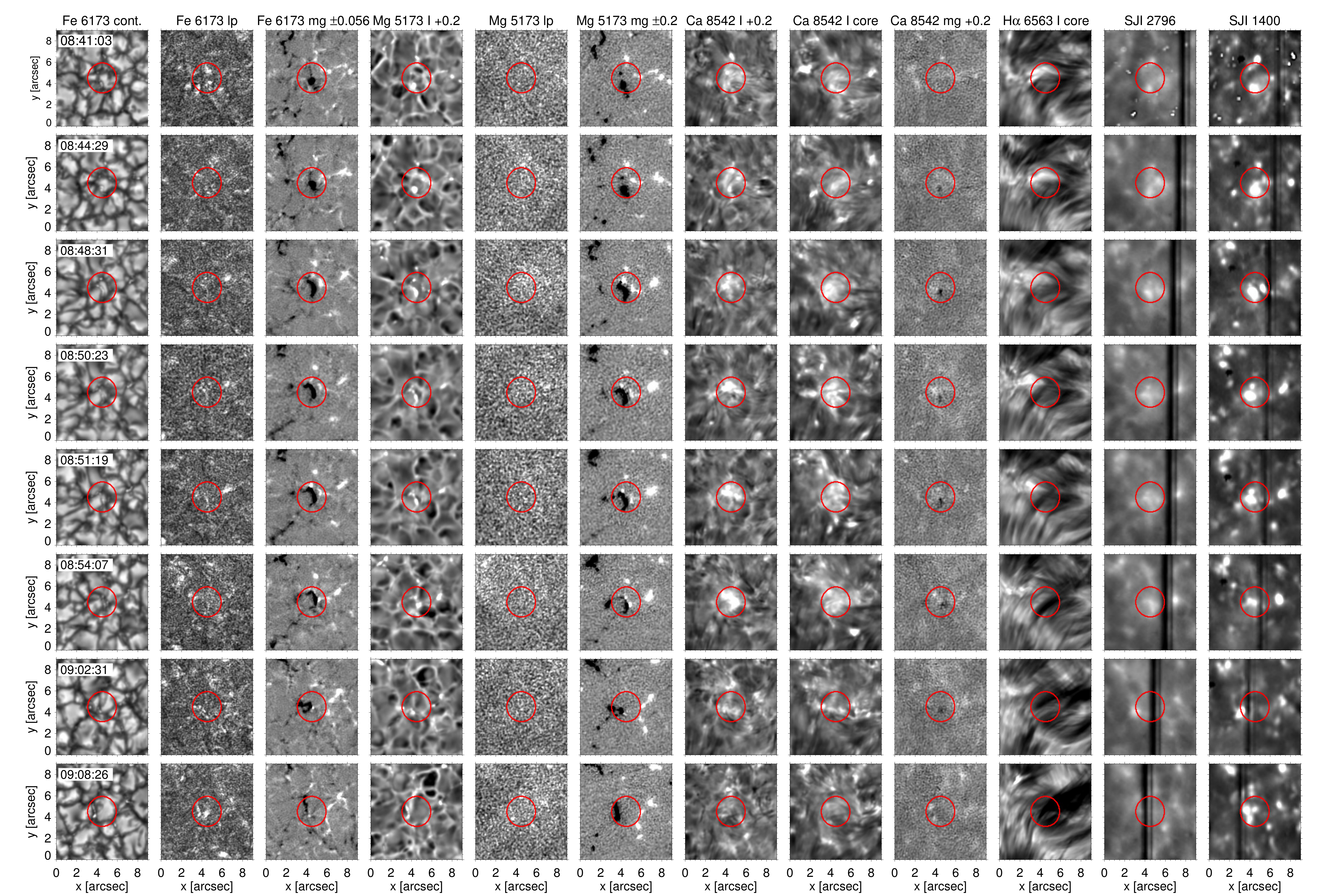}}
	\end{center}
	\vspace*{-1em}
	\caption{Same as Figure \ref{fig4} but for example 2 of IN flux cancellation. The temporal sequences show again, from left to right, continuum intensity, linear polarization and magnetograms in \ion{Fe}{1} 6173~\AA, intensity images in the red wing, linear polarization maps and magnetograms recorded in the Mg 5173~\AA\ line, filtergrams in the red wing and core of the \ion{Ca}{2} 8542~\AA\ line,  \ion{Ca}{2} magnetograms, the core in H$\alpha$ 6563~\AA, IRIS SJI 2796 and IRIS SJI 1400 filtergrams. The observed canceling IN event is enclosed by the red circle. IN flux concentrations located around $(x, y)=(5\arcsec, 5\arcsec)$ start interacting around 08:42:37~UT as bright features appear in the filtergrams showing the chromosphere and transition region. Fe and Mg magnetograms reveal cancellation of IN flux concentrations at the photospheric level.\newline
		{\em An animation of this figure is available in the online journal.}}
	\label{fig6}
\end{figure*}

The first spectral line is taken at 08:11:49~UT, just before the appearance of strong bright structures in the \ion{Ca}{2} filtergrams. As cancellation unfolds with time, the spectral profiles in the canceling region start to be distinctively different from the QS average. The line center clearly shifts toward the blue side of the line while developing strong asymmetry. There is a very sharp gradient on the red side and a much more extended blue side. The line intensity shows enhancement both in the wings and core during the cancellation. In addition to this, at $0.3$ and $0.4$~\AA\ the line clearly turns into emission, which is not present in the blue wing. We will discuss in Section \ref{sect5} a possible mechanism that may be responsible for producing the observed \ion{Ca}{2} $8542$~\AA\ Stokes $I$ profiles.

\subsubsection{Example \#2}

Example 2 is more involved but provides essential information on the cancellation process through available IRIS spectra. This example is displayed in Figure \ref{fig6} and shows interaction between two negative and two positive flux patches, as detected in Mg magnetograms. The temporal evolution of this canceling event and the atmospheric layers above can also be seen in the accompanying movie. It shows the same observables as in the case of example 1. 

From the beginning of the temporal subsequence, the SST continuum intensity maps expose bright points coinciding with the magnetic elements involved in cancellation (enclosed with red circles). At the same time, the negative polarity patches are co-spatial with localized brightenings, clearly visible in filtergrams taken in the wing and core of \ion{Ca}{2} $8542$~\AA. These brightenings appeared as soon as the positive flux patch at $(x, y)=(4\farcs5, 6\farcs5)$ approached the negative patches (this spatio-temporal evolution is not shown here). On the other hand, just like in the first example, there is a noticeable chromospheric activity between positive and negative flux concentrations appearing at 8:42:37~UT in the wing of \ion{Ca}{2} $8542$~\AA. Small bright patches appear above the positive magnetic elements at $(x, y)=(5\farcs5, 5\arcsec)$. This may indicate the onset of reconnection where associated currents are causing heating and produce brightenings along the field lines connecting the footpoints. In the next frame, a bright feature shows up in the core of the \ion{Ca}{2} $8542$~\AA\ line and is located between positive and negative flux structures. Only one minute later, at 8:43:51~UT, IRIS SJI 1400 filtergrams reveal signal intensification in the same region.

\begin{figure*}[t]
	\begin{center}
		\resizebox{0.5\hsize}{!}{\includegraphics[]{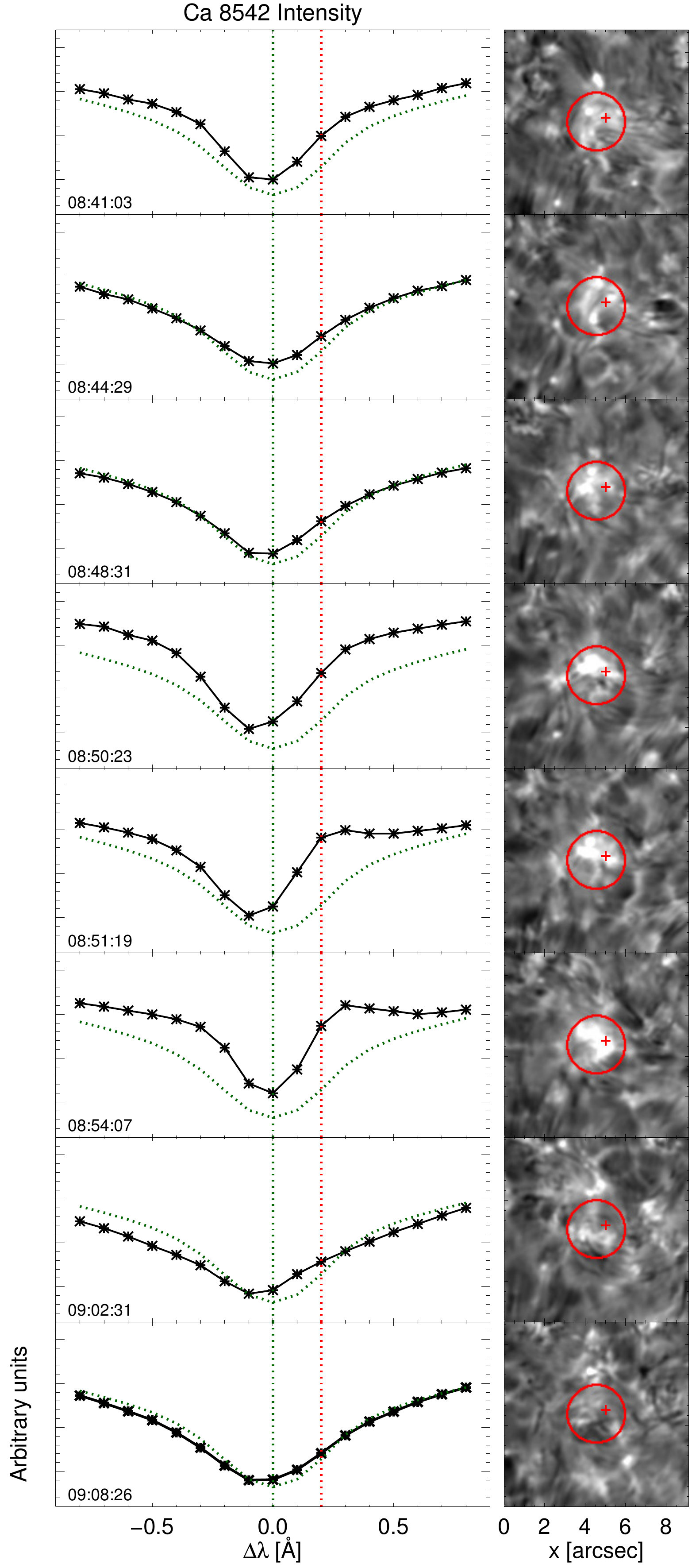}}
	\end{center}
	\vspace*{-1em}
	\caption{Same as Figure \ref{fig5} but for example 2. The temporal evolution of Stokes $I$ profiles in the \ion{Ca}{2} $8542$~\AA\ line (black lines) show the line intensity enhancement in the
wings and core during the cancellation while the red wing of the \ion{Ca}{2} $8542$~\AA\ line turns into emission.}
	\label{fig7}
\end{figure*}

\begin{figure}[t]
	\begin{center}
		\resizebox{1\hsize}{!}{\includegraphics[]{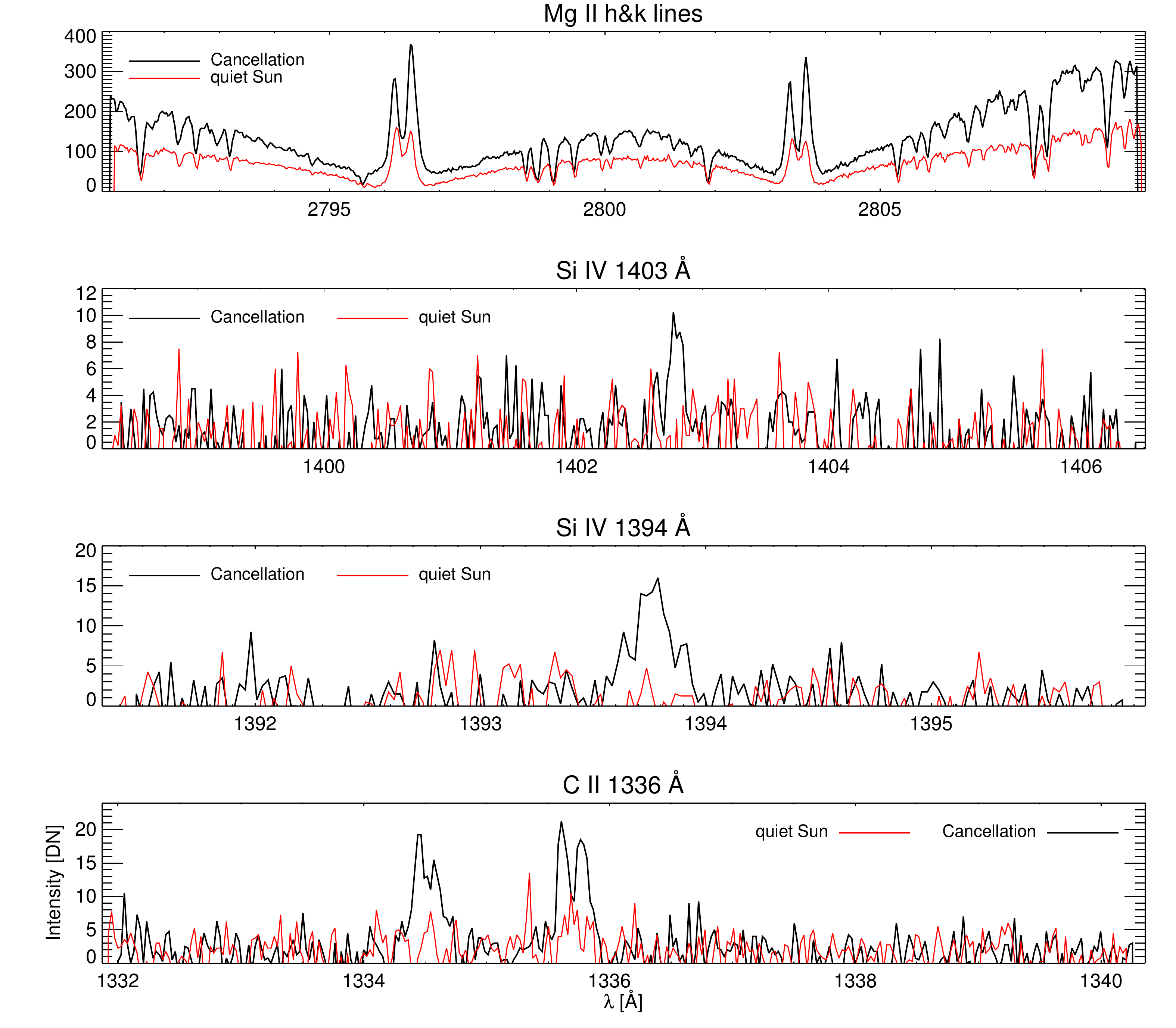}}
	\end{center}
	\vspace*{-1em}
	\caption{IRIS spectra recorded during the cancellation event presented in Figure \ref{fig6}. Panels, from top to bottom: the IRIS \ion{Mg}{2} h \& k lines with the surrounding spectral lines and continuum, the Si IV $1400$~\AA\ and $1394$~\AA\ lines, and C II $1336$~\AA\ spectral window. The black solid lines represent the IRIS spectra during the cancellation process, while the red line corresponds to the QS pixel outside of canceling region. The spectra are recorded at 8:55:59~UT.}
	\label{fig8}
\end{figure}

Magnetic flux cancellation is detected for the first time in Mg magnetograms at 8:44:47~UT when two small opposite polarity elements start interacting $(4\farcs5, 5\arcsec)$. Meanwhile, \ion{Ca}{2} filtergrams show the evolution of bright features above the canceling site. They seem to take the form of loop-like structures, oriented in a way to indicate connection between interacting magnetic elements. 

The other two opposite-polarity flux patches start canceling at 8:47:35~UT, as another roundish bright feature develops in IRIS SJI 1400, just left of the slit. While cancellation unfolds in the photosphere, H$\alpha$ measurements show the formation of dark fibrils at $(x, y)=(4\arcsec, 5\arcsec)$. By 8:56:55~UT, the smaller positive flux patch completely disappeared. During this period, the signal in the wing and core of \ion{Ca}{2} $8542$~\AA\ remained strong, showing dynamic bright structures above the canceling site. The dark fibrils visible in H$\alpha$ core grow with time and are oriented perpendicular to the polarity inversion line. Part of the fibrils dissolved around 9:00:39~UT and part got mixed with other overlying H$\alpha$ features.

At 9:03:27~UT the second positive polarity patch disappeared from Mg magnetograms, but can be observed further in Fe magnetograms, until the end of the temporal sequence. LP signal is visible only in the Fe $6173$~\AA\ line at the end of the movie and is probably not generated by the canceling event, but due to the flux emergence in the vicinity, at $(x, y)=(4\farcs5, 4\farcs5)$. Bright features disappeared from \ion{Ca}{2} wing filtergrams at 9:05:56~UT and from the core at 9:07:48~UT. 
 
The left column in Figure \ref{fig7} displays the temporal evolution of Stokes I profiles in the \ion{Ca}{2} $8542$~\AA\ line, formed inside the canceling region as denoted with the red cross. The corresponding \ion{Ca}{2} $8542+0.2$~\AA\ filtergrams are given in the right column. The intensity profiles exhibit very similar temporal evolution to those in example 1. Before and at the beginning of the cancellation process, intensity profiles are similar to that in the average QS, although slightly blue shifted. With time, the \ion{Ca}{2} $8542+0.2$~\AA\ line exhibits enhancement at all wavelengths and becomes asymmetric with the red wing (from $0.3$ to $0.4$~\AA) turning into emission around 8:51:19~UT. The profiles are again clearly shifted toward the blue side during cancellation. 

Characteristic IRIS spectra for case \#2 are given in Figure \ref{fig8}. The black solid line shows the spectra taken at the canceling site at 8:55:59~UT (marked with red cross in Figure \ref{fig7}). The non-magnetic QS spectra are shown with the red solid line. The \ion{Mg}{2} h \& k lines, which are formed above the flux cancellation site, are much stronger and more asymmetric than in the ``non-magnetic'' QS environment. There is a clear intensity enhancement throughout the \ion{Mg}{2} k and h spectral domain, where the k2r and h2r are stronger than k2v and h2v peaks. In addition to this, the continuum intensity also shows an increase during the flux cancellation. Such a profile may indicate a substantial amount of energy released through the cancellation process, resulting in local heating of the upper chromospheric layers. The intensity ratio between the blue and the red peak suggests a possible presence of upflowing plasma \citep{Leenaartsetal2013b}. Si IV $1400$~\AA, Si IV $1394$~\AA, and C II $1336$~\AA\ lines provide additional information on the upper chromospheric/transition region heating. These lines all show small, but noticeable intensity increases (black solid lines) compared to the non-magnetic QS profiles that are completely buried in the noise (red solid lines). 

\begin{figure*}[t]
	\begin{center}
		\resizebox{1\hsize}{!}{\includegraphics[]{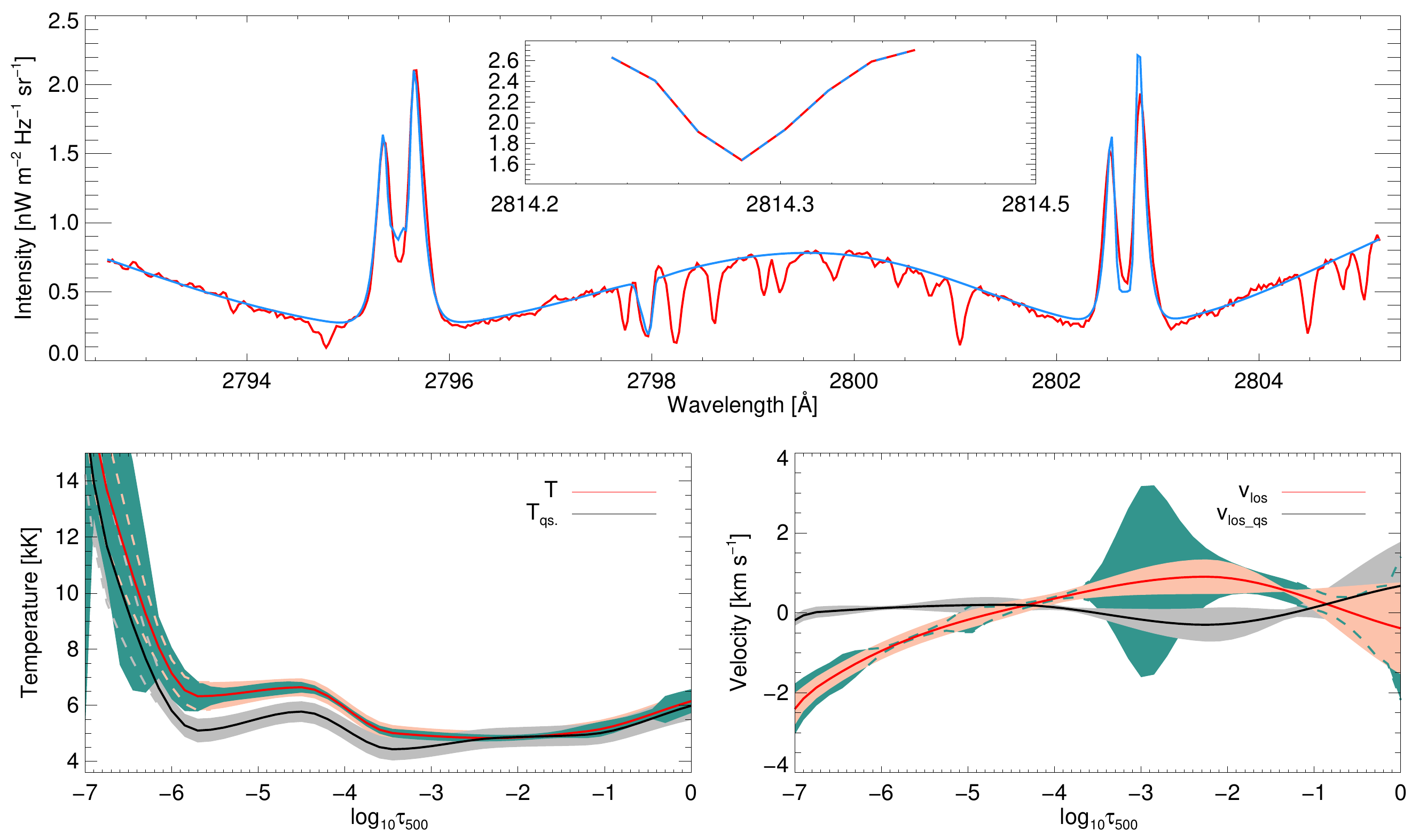}}
	\end{center}
	\vspace*{-1em}
	\caption{\textit{Upper panel}: Observed \ion{Mg}{2} h \& k spectra (red) and best-fit (blue) for a pixel inside the canceling region for which the IRIS \ion{Mg}{2} h \& k and
		UV triplet lines are inverted. The inset depicts the observed \ion{Ni}{1}~$2814.350$~\AA\ line (red) used to better constrain velocities in the photosphere and best-fit (dashed blue line). \textit{Bottom panels}: The retrieved temperature (left) and $v_{\text{los}}$ (right) are indicated with the red curves. Here we use the convention that positive values represent downflowing plasma. As a reference, the black lines show the inverted quiet Sun temperature and $v_{\text{los}}$ stratifications. Their respective uncertainties from multiple inversions are shown with the gray and red shaded areas. Those uncertainties calculated from the response functions are plotted with the green shaded areas.}
	\label{fig11}
\end{figure*}

\begin{figure*}[t]
	\begin{center}
		\resizebox{1\hsize}{!}{\includegraphics[]{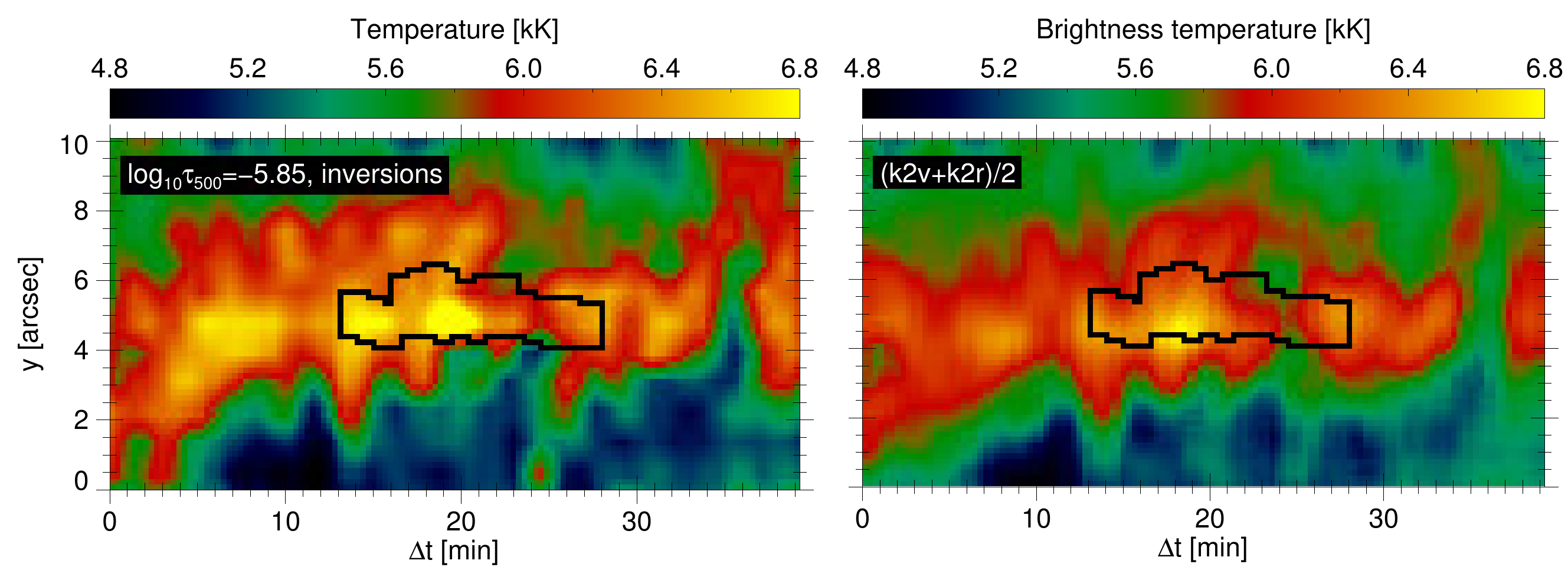}}
	\end{center}
	\vspace*{-1em}
	\caption{\textit{Left panel}: Temperature map at $\text{log}_{10}\tau_{500}=-5.85$ derived from the inversions of IRIS \ion{Mg}{2} h \& k lines in pixels at the canceling site (black contour) and its surroundings. \textit{Right panel}: The brightness temperature obtained from the average IRIS \ion{Mg}{2} k2v and k2r peak intensities. $\Delta t=0$~min corresponds to 08:41:03~UT.}
	\label{fig12}
\end{figure*}

In order to decipher whether the \ion{Mg}{2} h \& k profiles shown in Figure \ref{fig8} indicate heating of the upper solar atmosphere, we inverted these lines using STiC. The best-fit is presented with the blue solid line in the upper panel of Figure \ref{fig11}. The red solid curve shows the observed spectra. The \ion{Ni}{1} line used to better constrain the LOS velocities in the photosphere is shown in the inset. Here the best-fit is plotted with a dashed blue curve for the purpose of visualization. The observed profiles are very well reproduced, except for small discrepancies in the core of the \ion{Mg}{2} lines. The temperature stratification in the resulting model atmosphere is displayed in the bottom left panel of Figure \ref{fig11}. It can be seen that the photospheric and chromospheric temperatures inferred from the inversions in the case of cancellation (red line) are higher than in the QS case (black line). The corresponding shaded areas represent their respective uncertainties. Our results suggest that the temperature in the solar atmosphere increases by up to $2000$~K when internetwork magnetic elements start canceling. The temperature difference between the canceling and non-magnetic QS cases is the highest in the range $\text{log}_{10}\tau_{500}=[-6.5, -4]$. This is expected because the canceling \ion{Mg}{2} profiles normally has much broader wings and significantly stronger k and h peaks, which is mainly determined by the temperature stratification. Further evidence of the temperature increase around the canceling region is provided in Figure \ref{fig12} (left panel). The coarse map (every 3$^{\text{rd}}$ pixel along the slit per 3$^{\text{rd}}$ time step is selected) shows the temperature at $\text{log}_{10}\tau_{500}=-5.85$, where one of the temperature nodes is located (this corresponds to the mid-chromospheric layers in the FALC model). We can see the temperature distribution in the vicinity of the canceling region. The black contour indicates the borders of the canceling flux patches detected in the photosphere as the slit passes across them. Interestingly, the highest temperatures are found just above the canceling site. The reliability of the inversions and the reconstructed temperatures can also be assessed by comparing the temperature map inferred from the inversions (left panel) with the map constructed from the average of the IRIS \ion{Mg}{2} k2v and k2r peak intensities. Those \ion{Mg}{2} components are, based on comparisons with advanced numerical simulations, considered to be good indicators of chromospheric temperatures \citep{Pereiraetal2013}. The Pearson linear correlation coefficient for these two maps is $0.81$.

\begin{figure*}[t]
	\begin{center}
		\resizebox{1\hsize}{!}{\includegraphics[]{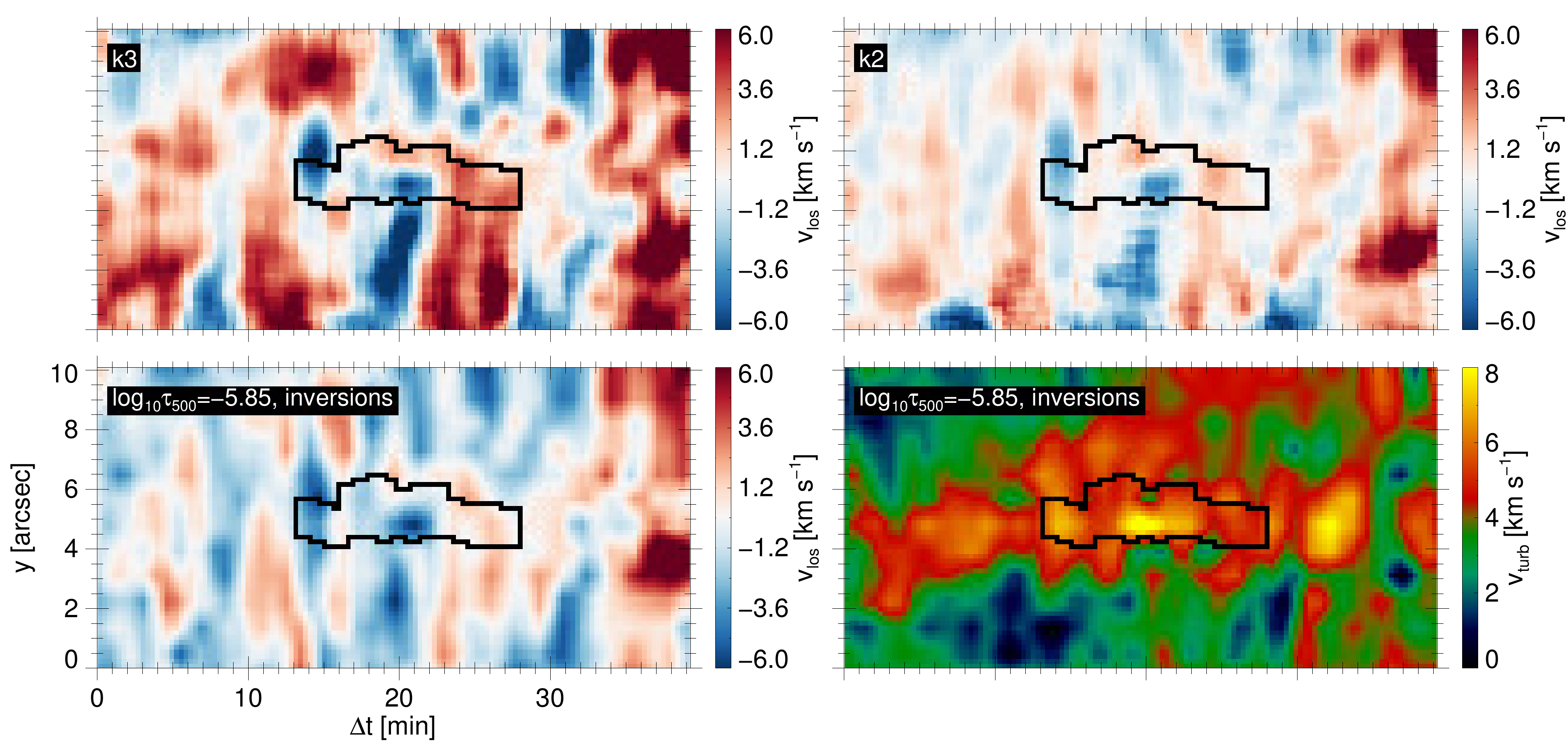}}
	\end{center}
	\vspace*{-1em}
	\caption{The reconstructed velocity maps associated with the temperature distribution from Figure \ref{fig12}. \textit{Upper left}: LOS velocities derived from the Doppler shift of the IRIS \ion{Mg}{2} k3 line core. \textit{Upper right}: LOS velocities obtained from the \ion{Mg}{2} k2 mean Doppler shifts. \textit{Lower left}: Reconstructed LOS map from inversions. \textit{Lower right}: Microturbulence velocities from inversions. The canceling region is enclosed by the black contour.}
	\label{fig13}
\end{figure*}

LOS velocities in the canceling and non-magnetic pixels are marked with solid red and black lines in the bottom right panel of Figure \ref{fig11}, respectively. Again, the shaded areas show their uncertainties. Since the \ion{Mg}{2} lines are insensitive to perturbations of the photospheric LOS velocities, this physical parameter is mostly constrained by the \ion{Ni}{1} line and hence less trustworthy at photospheric heights. In any case, the profiles mostly show either zero or positive values (i.e., downward) throughout the lower solar atmosphere. The canceling pixel shows upflows in the upper chromosphere, where IRIS \ion{Mg}{2} spectral lines are sensitive and provide more reliable values. On the other hand, $v_{\text{los}}$ in the QS pixel remains around zero. Figure \ref{fig13} reveals $v_{\text{los}}$ and microturbulence maps above the canceling site (enclosed by the dark contours) and its surroundings. The panels in the upper row show $v_{\text{los}}$ derived from the Doppler shifts of the IRIS \ion{Mg}{2} k$_{3}$ (left) and k$_{2}$ (right) components following the recipe by \cite{Pereiraetal2013}. We can see that these maps display practically the same features although the velocities are higher in the k$_{3}$ map. $v_{\text{los}}$ reconstructed from inversions is shown in the lower left panel and despite having smaller amplitudes, it resembles the k$_{2}$ map providing confidence in the inversion results. The Pearson linear correlation coefficient has its highest value of 0.72 at $\text{log}_{10}\tau_{500}=-5.85$. As can be seen, there are strong upflows inside the canceling region, reaching speeds of up to $-7$~km~s$^{-1}$ at height of $\text{log}_{10}\tau_{500}=-5.85$. These results indicate the presence of plasma being pushed up to the chromosphere at an average speed of $-4$~km~s$^{-1}$. It is reasonable to assume that this plasma, if it stays at the same temperature, should eventually fall down to the surface and downflows would be detected. However, we do not see this effect in our observations because the slit is moving across the canceling region and therefore, does not sample the same environment all the time. Still, we see downflows in several pixels closer to the end of the canceling event, from $\Delta t=23$~min until $\Delta t=27$~min. Reconnection of magnetic field lines increases the temperature locally, but seems to also cause higher microturbulent velocities in the chromosphere (lower right panel in Figure \ref{fig13}). The similarity of the microturbulence and temperature maps may also be, to some extent, the result of an interplay between these two atmospheric parameters since the wing widths of the \ion{Mg}{2} lines and the subordinate triplet lines are sensitive to both of them (\citealt{Carlssonetal2015}, see also Figure 5 in \citealt{delaCruzRodriguezetal2016}).

\subsection{Energy budget of IN cancellations}

Our observations suggest that the process of IN flux cancellation provides a substantial amount of energy that is capable of heating the chromosphere locally. In what follows we estimate the energy content of IN canceling magnetic patches.

\subsubsection{Magnetic energy}

Based on our tracking results, i.e., considering the number of detected canceling patches, their flux content, and the total FOV,	the flux cancellation rate is $27$~Mx~cm$^{-2}$~day$^{-1}$, which is in agreement with the cancellation rate reported by \cite{Gosicetal2016}.

\begin{figure*}[t]
	\begin{center}
		\resizebox{1\hsize}{!}{\includegraphics[]{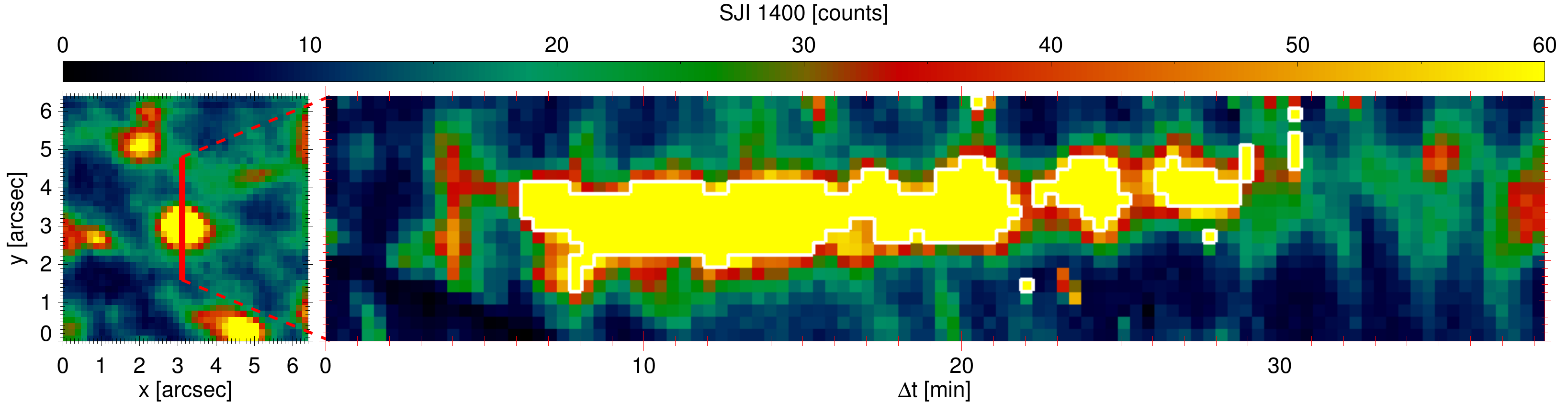}}
	\end{center}
	\vspace*{-1em}
	\caption{Time slice across the canceling event shown in Figure \ref{fig4}. The image on the left highlights the detected SJI 1400 bright grain above the canceling region with its surroundings at $\Delta t$=19.6~min. The red solid vertical line corresponds to the cuts used for the time slice image displayed in the right panel. The white contour coincides with the threshold of 60 counts used to detect bright structures in IRIS SJI 1400 filtergrams.}
	\label{fig17}
\end{figure*}

To derive the magnetic field strength ($B$) from the observed Stokes profiles in the Mg 5173~\AA\, we carried out inversions using the SIR code \citep{SIR1992}. This code numerically solves the radiative transfer equation under the assumption of local thermodynamic equilibrium (LTE) and provides the temperature stratification, the velocity, the magnetic field strength, inclination and azimuth angles along the line of sight. Inverting a non-LTE line with a LTE code means that the retrieved thermal parameters are not trustable, but the derived magnetic field stratification can be considered reliable (see Fig. 14 in \citealt{delaCruzRodriguezetal2012}). We used three nodes in the temperature while the magnetic field strength and LOS velocity are assumed to have a linear gradient with height, which is necessary in order to fit the Stokes V profiles that show an asymmetry between the blue and red lobes \citep[for a detailed review see][]{DelToroIniestaRuizCobo}. The rest of the atmospheric parameters are kept constant with height. In order to reduce the number of free parameters, we assumed the magnetic filling factor to be equal to one and did not use stray light contamination, which are reasonable assumptions when high resolution observations are inverted as in this case. We use the Harvard Smithsonian Reference Atmosphere \citep{Gingerichetal1971} as the initial model atmosphere. The average magnetic field strength of canceling IN patches (at the moment when cancellation starts), is found to be about $160$~G, which is roughly in agreement with the average strength of IN fields reported earlier \citep[e.g., 220~G;][]{OrozcoBellotRubio2012}. This allows us to estimate the magnetic energy density per canceling event:

\begin{equation}
E_{mag}=\frac{B^{2}}{8\pi}\approx1020\ \text{erg}~\text{cm}^{-3}.
\label{eq2}
\end{equation}

Considering that IN cancellations last about $t=3.6$ minutes at the photospheric level, we obtain an energy release rate of $\Delta E_{mag}\approx4.7$~erg~cm$^{-3}$~s$^{-1}$. This shows that the energy accumulated inside individual canceling regions is sufficient to locally balance the radiation loss of $10^{-1}$~erg~cm$^{-3}$~s$^{-1}$ in the lower chromosphere \citep{Vernazzaetal1981}.

However, taking into account the occurrence rate of canceling events ($n=411$) inside the observed FOV ($S_{\text{FOV}}=1.2\times10^{19}$~cm$^{2}$), the horizontal extent of the reconnecting regions ($S_{\text{r}}=0.6$~arcsec$^2$), and the duration of our observations ($T=2.6$~hr), we can see that the energy contribution from IN cancellations averaged over time and area is $\Delta E_{mag}nS_{\text{r}}t/S_{\text{FOV}}T\approx1\times10^{-2}$~erg~cm$^{-3}$~s$^{-1}$. In other words, the total energy input is an order of magnitude lower than what is necessary to globally maintain the chromospheric heating.

We want to stress here that our IRIS and SST observations clearly show that IN cancellations are local phenomena. This is also demonstrated with the time slice image shown in the right panel of Figure \ref{fig17}. As can be seen, the signal in IRIS SJI 1400 images is highly localized, and beyond the threshold of 60 counts (outlined by the white contour), drops rapidly to the background level. This indicates that the energy flux mostly propagates vertically, and does not significantly affect the solar atmosphere in horizontal directions. Because of this, and their small filling factor, the IN canceling events that we study here (i.e., at the sensitivity levels of our current dataset), seem to play a minor role in globally heating the internetwork chromosphere, despite of the high local excess of energy flux.

\subsubsection{Missing magnetic energy}

We have established that the observed IN cancellations produce significant temperature increases in the chromosphere, but only locally. To play a global role in chromospheric heating, a much larger area coverage would certainly be needed.

With the current data we cannot exclude that cancellations occur ubiquitously, however. First, the sensitivity of our magnetograms is of the order of 24 G (3 times the sigma level). In principle, it is possible that weaker fields are buried in the noise. Actually, very deep observations with the Hinode Spectro-Polarimeter \citep[SP;][]{Litesetal2013} have shown that the IN is nearly fully covered by weak fields \citep{OrozcoBellotRubio2012, BellotRubioOrozcoSuarez2012}. If fields of $\sim20$~Mx~cm$^{-2}$ are present all over the solar surface - as seems to be the case - and dissipate over the same time scales as the IN cancellations we observe with the SST, then they could explain the global heating of the IN chromosphere.

The challenge is thus to increase the sensitivity of the observations at least up to the level of the very deep Hinode/SP measurements and, at the same time, maintain high temporal resolution in order to detect the weakest IN fields, increase the surface area coverage, and investigate if they cancel out in a similar way as the stronger elements visible with the SST.

Only magnetic field measurements that are more sensitive can fully address the role of IN cancellations in the global heating of the IN chromosphere. The IRIS and SST observations analyzed here suggest that their area coverage is too small to provide the energy needed to sustain the chromosphere, but future measurements with the Daniel K. Inouye Solar Telescope \citep[DKIST;][]{Elmoreetal2014} and the European Solar Telescope \citep[EST;][]{Colladosetal2013}, will allow more definitive statements to be made.

\section{Discussion and conclusions}
\label{sect5}

In this paper we studied cancellations of IN magnetic elements in an attempt to decipher their impact on the chromospheric and coronal heating. To this purpose, we carried out analysis of simultaneous SST and IRIS observations of IN regions. Using SST magnetograms in the Mg 5173~\AA\ line we identified all IN flux features that undergo either partial or total cancellations in the photosphere. Thanks to multi-wavelength measurements, we were able to observe the temporal evolution of these magnetic elements and their coupling to the upper solar atmosphere. In particular, we analyzed SST and IRIS spectral line profiles at the locations where IN flux cancellations occur, sampling the solar atmosphere from the photosphere up to the transition regions.

Using SST \ion{Mg}{1} b$_2$ magnetograms we identified 411 canceling events distributed all over the FOV. Most of them ($360$) involve IN flux concentrations and are covered by the IRIS FOV. In total, 76\% of the detected events are cospatial with bright grains identified in IRIS SJI 1400 images, suggesting possible heating in the chromosphere. On average, cancellations seen in Mg magnetograms have shorter lifetimes than SJI 1400 bright structures, $\sim$$3$~min compared to $\sim$$12$~min, respectively. This result indicates that SJI 1400 bright grains above canceling regions, may be the result of reconnection of internetwork magnetic field lines in the chromosphere (and above) that starts earlier than the flux cancellations in the photosphere.

In fact, interpretation of the photospheric flux cancellations as a consequence of magnetic reconnection in the chromosphere has been reported by multiple studies \citep[e.g.,][]{Chaeetal1998, Chaeetal1999, Chaeetal2010, Wangetal2000}. The temporal evolution of the canceling regions analyzed in this work, implies that chromospheric reconnection is a plausible scenario for the quiet Sun IN cancellations. When two or more opposite polarity patches, that were not connected before, are approaching toward each other, magnetic field lines start to reconnect at the chromospheric level. As a consequence, the released energy produces a local brightening in the chromosphere and possibly in the transition region. Such imprints are seen in both our examples. In particular, we followed opposite polarity flux patches as they approach. Before they start canceling in the photosphere, there are strong brightenings in the chromosphere as seen in \ion{Ca}{2} 8542~\AA\ intensity maps. They are located exactly above canceling regions.

The spectral profiles of the \ion{Ca}{2} 8542~\AA\ line have a very characteristic asymmetric form, clearly shifted toward the blue with the red wing in emission. Such line shape has been reported to be associated with regions where strong local upflows are present and may be coupled with temperature enhancements caused by magnetic reconnections somewhere in the atmosphere \citep{Ortizetal2014, delaCruzRodriguezetal2015a}. We obtained more evidence for the upflowing plasma inside canceling regions from the properties of IRIS \ion{Mg}{2} h \& k lines, where according to \cite{Leenaartsetal2013b}, a stronger k2r peak implies upflow in the upper chromosphere. This is consistent with our $v_{\text{los}}$ derived from the non-LTE inversions of the \ion{Mg}{2} h \& k and UV triplet lines, shifts of the IRIS \ion{Mg}{2} k3 line cores, and from the \ion{Mg}{2} k2 mean Doppler shifts. We found that upflow velocities dominate inside canceling regions. We also detected bright structures in IRIS SJI 1400 filtergrams which supports the possibility of heating in the middle chromosphere and transition region since those filtergrams sample the region from the upper photosphere up to the transition region \citep{MartinezSykoraetal2015}. During the later phase of flux cancellation, dark fibrils appear in the core of the H$\alpha$ line, which is likely a signature of plasma ejected upward by reconnection processes \citep{Chaeetal2010}. 

We have demonstrated that the magnetic energy density inside canceling regions is sufficient to produce a local temperature increase in the chromosphere by more than $10^{3}$~K. However, the area covered by cancellation events is not high enough to globally support the radiative losses necessary for the chromospheric heating \citep[$10^{-1}$~erg~cm$^{-3}$~s$^{-1}$;][]{Vernazzaetal1981}. It is possible that numerous canceling magnetic elements are not visible in our observations due to limited spatial resolution and sensitivity. These fields would be too small and/or weak to produce polarization signals above the noise level. In such a case, we would lose their contribution and our calculated magnetic energies would be underestimated. To investigate the possible existence of IN fields on sub-resolution scales that contribute to chromospheric heating we need measurements exceeding the capabilities of existing telescopes. The upcoming 4m solar telescopes such as DKIST and EST will reach sensitivities of $10^{-4}$ and spatial resolutions of some 20 km with exposure times of a few seconds only. With such a sensitivity we may be able to detect the weakest IN fields, while the unprecedented spatial resolution will help us detect fields below the currently accessible spatial and temporal scales.

In the meantime, according to our results, cancellations of IN flux patches play a role in the local energy balance in and around the canceling sites, but cannot maintain global chromospheric heating by themselves, supporting the results of \cite{Wiegelmannetal2013} and \cite{Chittaetal2014}. However, IN cancellations contribute along the other possible heating mechanisms. One such mechanism, for example, could be heating of the upper solar atmosphere by emergence of magnetic flux and rise of the associated magnetic field lines up to the chromosphere and the transition region. This phenomenon will be addressed in our future work.  

\acknowledgments 
IRIS is a NASA small explorer mission developed and operated by LMSAL with mission operations executed at NASA Ames Research center and major contributions to downlink communications funded by ESA and the Norwegian Space Centre. MG was supported by NASA grant NNX16AC34G. JdlCR is supported by grants from the Swedish Research Council (2015-03994), the Swedish National Space Board (128/15) and the Swedish Civil Contingencies Agency (MSB). This project has received funding from the European Research Council (ERC) under the European Union's Horizon 2020 research and innovation programme (SUNMAG, grant agreement 759548). BDP was supported by NASA grant NNX11AN98G and NASA contracts NNG09FA40C (IRIS). The work of LBR and SEP was supported by the Spanish Ministerio de Econom{\'i}a and Competitividad through grants ESP2013-47349-C6-1-R and ESP2016-77548-C5-1-R, including a percentage from European FEDER funds. Image reconstruction was performed at IAA-CSIC supercomputing facilities. The Swedish 1~m Solar Telescope is operated by the Institute for Solar Physics of Stockholm University in the Spanish Observatorio del Roque de los Muchachos of the Instituto de Astrof\'isica de Canarias. MC was supported by the Research Council of Norway through its Centres of Excellence scheme, project number 262622, and through grants of computing time from the Programme for Supercomputing. This research has made use of NASA's Astrophysics Data System.

\end{document}